\documentclass[preprint,
amsmath,amssymb,aps, nofootinbib]{revtex4-1}

\usepackage{graphicx}
\usepackage{dcolumn}
\usepackage{bm}
\usepackage{color}

\usepackage{geometry} 
\usepackage{fancyhdr} 
\usepackage{moresize} 

\usepackage{multirow}
\usepackage{longtable}
\usepackage{xcolor}
\usepackage{colortbl,booktabs}

\newcommand{\la}{\langle}
\newcommand{\ra}{\rangle}
\newcommand{\qbar}{\bar{q}}
\newcommand{\ubar}{\bar{u}}

\newcommand{\sbar}{\overline{s}}
\newcommand{\cbar}{\overline{c}}

\def\non{\nonumber}
\def\up{\uparrow}
\def\dw{\downarrow}
\def\be{\begin{eqnarray}}
\def\en{\end{eqnarray}}

\begin{document}


\title{Two-body hadronic weak decays\\ of antitriplet charmed baryons}

\author{Jinqi Zou, Fanrong Xu\footnote{fanrongxu@jnu.edu.cn},
Guanbao Meng}
\affiliation{
 Department of Physics, Jinan University,
 Guangzhou 510632, People's Republic of China
}%

\author{Hai-Yang Cheng}
\affiliation{%
 Institute of Physics, Academia Sinica, Taipei, Taiwan 115, Republic of China }
 %

\bigskip
\begin{abstract}
\bigskip
We study Cabibbo-favored (CF) and singly Cabibbo-suppressed (SCS) two-body hadronic weak decays of the antitriplet charmed baryons $\Lambda_c^+$, $\Xi_c^0$ and $\Xi_c^+$ with more focus on the last two. Both factorizable and nonfactorizable contributions are considered in the topologic diagram approach.
The estimation of nonfactorizable contributions from $W$-exchange and inner $W$-emission diagrams relies on the pole model and current algebra. The non-perturbative parameters in both factorizable and nonfactorizable parts are calculated in the MIT bag model.
Branching fractions and up-down decay asymmetries for all the CF and SCS decays of antitriplet charmed baryons are presented.
The prediction of $\mathcal{B}(\Xi_c^+\to
\Xi^0\pi^+ )$ agrees well with the measurements inferred from Belle and CLEO, while
the calculated ${\cal B}(\Xi_c^0\to \Xi^-\pi^+)$ is too large compared to the recent Belle
measurement. We conclude that these two $\Xi_c\to\Xi \pi^+$ modes cannot be simultaneously explained within the current-algebra framework for $S$-wave amplitudes. This issue needs to be resolved in future study.
The long-standing puzzle with the branching fraction and decay asymmetry of $\Lambda_c^+\to\Xi^0 K^+$ is resolved by noting that only type-II $W$-exchange diagram will contribute to this mode. We find that not only the calculated rate agrees with experiment but also the predicted decay asymmetry is consistent with the SU(3)-flavor symmetry approach  in sign and magnitude.
Likewise, the CF mode $\Xi_c^0\to \Sigma^+K^-$ and the SCS decays $\Xi_c^0\to pK^-,\Sigma^+\pi^-$ proceed only through type-II $W$-exchange.  They are predicted to have large and {\it positive} decay asymmetries. These features can be tested in the near future.

\end{abstract}

\maketitle
\small


\section{Introduction}\label{sec:Intro}


Recently, there has been a significant progress in the experimental study of charm physics.
In the meson sector, LHCb measured the {\it CP} asymmetry difference between $D^0\to K^- K^+$ and $D^0 \to \pi^- \pi^+$,
giving $\Delta A_{\rm{CP}}=(-15.4\pm 2.9)\times 10^{-4}$ \cite{Aaij:2019kcg}, which is
the first observation of {\it CP} violation in the charm sector.
The  progress in charmed baryon physics is also impressive.
The long-quested doubly charmed baryon 
was first observed through
the process $\Xi_{cc}^{++}\to\Lambda_c^+ K^- \pi^+\pi^+$  at LHCb in 2017
\cite{Aaij:2017ueg}. Later in 2018, the lifetime of $\Xi_{cc}^{++}$
\cite{Aaij:2018wzf}, its mass and the two-body weak decay channel
$\Xi_{cc}^{++}\to\Xi_c^+\pi^+$
\cite{Aaij:2018gfl} were measured by LHCb.
Some breakthrough has also been made
in singly charmed baryons as well, especially the lightest one $\Lambda_c^+$.
Both Belle \cite{Zupanc} and BESIII \cite{Ablikim:2015flg}  have measured
the absolute branching fraction of the decay $\Lambda_c^+\to pK^-\pi^+$.
A new average of $(6.28\pm0.32)\%$ for this benchmark mode is quoted by the Particle Data Group (PDG)  \cite{Tanabashi:2018oca}.
The measurement of
$\Lambda_c^+\to p\pi^0, p\eta$ \cite{Ablikim:2017ors} indicated that singly Cabibbo-suppressed (SCS) decays are ready to access.

In addition to $\Lambda_c^+$, there have been some new developments  in the study of $\Xi_c^0$  and $\Xi_c^+$, the other two singly charmed baryons in the antitriplet.
By using a data set comprising $(772\pm11)\times 10^6$ $B\bar{B}$ pairs collected at
$\Upsilon(4S)$ resonance, Belle was able to measure the absolute branching fraction for $B^-\to \bar{\Lambda}_c^- \Xi_c^0$ \cite{Li:2018qak}.
Combining the subsequently measured product branching fractions such as $\mathcal{B}(B^-\to \bar{\Lambda}_c^- \Xi_c^0) \mathcal{B}(\Xi_c^0\to \Xi^-\pi^+)$, Belle reported the
first weak decay of $\Xi_c^0$ \cite{Li:2018qak},
\begin{equation}
\mathcal{B}(\Xi_c^0\to \Xi^-\pi^+)=(1.80\pm 0.50\pm 0.14)\times 10^{-2}.
\end{equation}
Using the same technique,  a channel of two-body weak decay with a vector meson in final state was
also measured, $\mathcal{B}(\Xi_c^+\to p \bar K^{0*}(892))=(0.25\pm 0.16\pm 0.04)\times 10^{-2}$ \cite{Li:2019atu}.
It is worth pointing out that the absolute branching fraction for three-body decay was  obtained by Belle \cite{Li:2019atu} to be
$\mathcal{B}(\Xi_c^+\to \Xi^-\pi^+\pi^+)=(2.86\pm 1.21\pm 0.38)\times 10^{-2}$, from which
we can read
\begin{equation}
\mathcal{B}(\Xi_c^+\to \Xi^0\pi^+)=
(1.57\pm0.83)\%,
\end{equation}
where use of
$\Gamma(\Xi_c^+\to \Xi^0\pi^+)/\Gamma(\Xi_c^+\to\Xi^-\pi^+ \pi^+)=
(0.55\pm0.13\pm0.09)$  obtained by the CLEO \cite{Edwards:1995xw} has been made.

\begin{table}[t]
\scriptsize{
 \caption{Branching fractions (upper entry) and up-down decay asymmetries $\alpha$ (lower entry) of Cabibbo-allowed $\Xi_c^{+,0}\to{\cal
B}+P$ decays in various early model calculations.  All the model results for branching fractions (in percent)
have been normalized using the current world averages of $\tau(\Xi_c^+)$ and $\tau_(\Xi_c^0)$ (see Eq. (\ref{eq:lifetime}) below).
 } \label{tab:CF-his}
\centering
\begin{ruledtabular}
\begin{tabular}{l c c c c c c c c}
~~~~Decay & K\"{o}rner,           & Xu,
& Cheng,             & \,\,\, Ivanov et al \, & \.Zenczykowski & Sharma,  & Expt. \\
        & Kr\"{a}mer \cite{Korner:1992wi} & Kamal \cite{Xu:1992vc}
& Tseng \cite{Cheng:1993gf} & \cite{Ivanov:1997ra} & \cite{Zenczykowski:1993jm} & Verma \cite{Sharma:1998rd}& \cite{Tanabashi:2018oca,Li:2018qak} \\
& & & CA \quad~ Pole & & &  & \\
 \hline
$\Xi_c^+\to \Sigma^+ \bar K^0$ & 6.66  & 0.46 & 0.05 \quad~ 0.87 & 4.05  & \\
$\Xi_c^+\to \Xi^0 \pi^+ $ & 3.65 & 3.47 & 0.87 \quad~ 4.06 & 5.78  & & & $1.57\pm0.83$ \\
$\Xi_c^0\to \Lambda \bar K^0$ & 0.17 & 0.50 & 1.36 \quad~ 0.37 & 0.55 &  \\
$\Xi_c^0\to \Sigma^0 \bar K^0$ & 1.61 & 0.14 & 0.03 \quad~ 0.18 & 0.26  &  \\
$\Xi_c^0\to \Sigma^+ K^-$ & 0.17 &0.17 & & 0.35  &  \\
$\Xi_c^0\to \Xi^0 \pi^0$ & 0.05 &0.77 & 1.71 \quad~ 0.38 & 0.05  &  \\
$\Xi_c^0\to \Xi^- \pi^+$ & 1.42 & 2.37 &1.13 \quad~ 1.71 & 1.60  & & & $1.80\pm 0.52$  \\
$\Xi_c^0\to \Xi^0 \eta$ & 0.32 & & & 0.37  &  \\
$\Xi_c^0\to \Xi^0 \eta'$ & 1.16 & & & 0.41  &  \\
\hline
$\Xi_c^+\to \Sigma^+ \bar K^0$ & $-$1.0 & 0.24 & 0.43 \quad
$-$0.09 &  $-$0.99 & 1.0&  0.54 & \\
$\Xi_c^+\to \Xi^0 \pi^+ $ & $-$0.78 & $-$0.81 & $-$0.77 \quad $-$0.77 & $-$0.97 & 1.0 & $-$0.27 &  \\
$\Xi_c^0\to \Lambda \bar K^0$ & $-$0.76 & 1.00 & $-$0.88 \quad $-$0.73 & $-$0.75 & $-$0.29 & $-$0.79 & \\
$\Xi_c^0\to \Sigma^0 \bar K^0$ & $-$0.96 & $-$0.99 &  0.85 \quad $-$0.59 & $-$0.55
& $-$0.50 & 0.48  &    \\
$\Xi_c^0\to \Sigma^+ K^-$ & 0 & 0 & & 0 & 0 & 0 &  \\
$\Xi_c^0\to \Xi^0 \pi^0$ & 0.92 & 0.92 &  $-$0.78 \quad $-$0.54 & 0.94 & 0.21 & $-$0.80 &  \\
$\Xi_c^0\to \Xi^- \pi^+$ & $-$0.38 & $-$0.38 & $-$0.47 \quad $-$0.99 & $-$0.84 & $-$0.79 & $-$0.97  & $-0.6\pm 0.4$ \\
$\Xi_c^0\to \Xi^0 \eta$ & $-$0.92 &  & & $-$1.0 & 0.21 &  $-$0.37 &  \\
$\Xi_c^0\to \Xi^0 \eta'$ & $-$0.38 & & & $-$0.32 & $-$0.04 &  0.56 &  \\
\end{tabular}
\end{ruledtabular}
 }
\end{table}

For lifetimes of the antitriplet  charmed
baryons, we quote the new world averages (in units of $10^{-13}$ s)
\begin{align} \label{eq:lifetime}
\tau(\Lambda_c^+)=2.03\pm0.02, \qquad
\tau(\Xi_c^+)=4.56\pm0.05, \qquad
\tau(\Xi_c^0)=1.53\pm0.02,
\end{align}
dominated by the most recent lifetime measurements by the LHCb \cite{Aaij:2019lwg}. Note that the measured $\Xi_c^0$ lifetime by the LHCb is approximately 3.3 standard deviations larger than the old world average value \cite{Tanabashi:2018oca}.

Inspired by latest experimental results of $\Xi_c$ decays,
there have been some efforts from theorists \cite{Geng:2018plk,Geng:2018bow,Geng:2019xbo,Zhao:2018mov,Zhao:2018zcb,Jia}.
Indeed, the study of charmed baryon weak decays,
including the charged and neutral $\Xi_c$ baryons, is an old subject.
To understand the underlying dynamical mechanism in hadronic weak decays, one may draw the topological diagrams according to
the hadron's content \cite{Chau:1995gk}.
In charmed baryon decays, nonfactorizable contributions from $W$-exchange or inner $W$-emission diagrams play an essential role and they cannot be neglected, in contrast with the negligible effects in heavy meson decays.
The fact that
all the decays of $\Xi_c^{+,0}$ receive nonfactorizable contributions, especially some decays
such as $\Xi_c^0\to\Sigma^+ K^-, \Xi^0\pi^0$
proceed only through  purely nonfactorizable diagrams,
will allow us to check the importance and necessity of nonfactorizable contributions.
However, we still do not have a reliable phenomenological model to
calculate charmed baryon hadronic decays so far.
In the 1990s various techniques were developed, including relativistic quark model (RQM) \cite{Korner:1992wi,Ivanov:1997ra}, pole model \cite{Xu:1992vc,Cheng:1993gf,Zenczykowski:1993jm} and
current algebra \cite{Cheng:1993gf,Sharma:1998rd},
to estimate the nonfactorizable effects in Cabbibo-favored $\Xi_c^{+,0}$ decays. The predicted branching fractions and decay asymmetries in various early model calculations are summarized in Table \ref{tab:CF-his}.
\footnote{For early model calculations of Cabibbo-allowed $\Lambda_c^+\to {\cal B}+P$ decays, see Table I of \cite{Cheng:2018hwl}.}

Now with more experimental data accumulated, there are some updated studies in theory  \cite{Geng:2018plk,Geng:2018bow,Geng:2019xbo,Zhao:2018mov,Zhao:2018zcb}.
In these works except \cite{Zhao:2018zcb}, the experimental results are taken as input and  global fitting analyses are carried out at the hadron level based on SU(3) flavor symmetry without resorting to the detailed dynamics.
Apparently, a reconsideration of charmed baryon weak decays, revealing the dynamics at the quark level, is timely and necessary.  Pole model is one of the choices.

In the pole model, important low-lying $1/2^+$
and $1/2^-$ states are usually considered under the pole approximation.
In the decay with a pseudoscalar in the final state, $\mathcal{B}_c\to\mathcal{B}+P$, the nonfactorizable  $S$- and $P$-wave amplitudes are dominated by $1/2^-$ low-lying baryon resonances and $1/2^+$ ground state baryons, respectively.
The $S$-wave amplitude can be further reduced to current algebra in the soft-pseudoscalar limit. That is, the evaluation of the $S$-wave amplitude does not require the information of the troublesome negative-parity baryon resonances which are not well understood in the quark model.
The methodology was developed and applied in the earlier work \cite{Cheng:1993gf}. It turns out if the $S$-wave amplitude is evaluated in the pole model or in the covariant quark model and its variant, the decay asymmetries for both $\Lambda_c^+\to \Sigma^+\pi^0$ and $\Sigma^0\pi^+$ were always predicted to be positive, while it was measured to be $-0.45\pm0.31\pm0.06$ for $\Sigma^+\pi^0$ by CLEO \cite{CLEO:alpha}. In contrast, current algebra always leads to a negative decay asymmetry for aforementioned two modes: $-0.49$ in \cite{Cheng:1993gf}, $-0.31$ in \cite{Sharma:1998rd}, $-0.76$ in \cite{Zenczykowski:1993hw} and $-0.47$ in \cite{Datta}.  The issue with the sign of $\alpha(\Lambda_c^+\to\Sigma^+\pi^0)$ was finally resolved by BESIII. The decay asymmetry parameters of $\Lambda_c^+\to \Lambda\pi^+,\Sigma^0\pi^+,\Sigma^+\pi^0$ and $pK_S$ were recently measured by BESIII \cite{Ablikim:2019zwe} (see Table \ref{tab:LambdacCF} below), for example,  $\alpha(\Lambda_c^+\to\Sigma^+\pi^0)=-0.57\pm0.12$ was obtained. Hence,
the negative sign of $\alpha(\Lambda_c^+\to\Sigma^+\pi^0)$ measured by CLEO is nicely confirmed by BESIII.
This is one of the strong reasons why we adapt current algebra to work out parity-violating amplitudes.

It is well known that there is a long-standing puzzle with the branching fraction and decay asymmetry of $\Lambda_c^+\to\Xi^0 K^+$. The calculated branching fraction turns out to be too small compared to experiment and the decay asymmetry is predicted to be zero owing to the vanishing $S$-wave amplitude. We shall examine this issue in this work and point out a solution to this puzzle. This has important implications to the $\Xi_c^0$ sector where
the CF mode $\Xi_c^0\to \Sigma^+K^-$ and the SCS decays $\Xi_c^0\to pK^-,\Sigma^+\pi^-$ will encounter similar problems in naive calculations.

Recently, we have followed this approach to
calculate singly Cabibbo-suppressed (SCS) decays of $\Lambda_c^+$ \cite{Cheng:2018hwl},
in which the predictions of $\Lambda_c^+\to p\pi^0, p \eta$ are in good agreement with the BESIII measurement.  In this work,
we shall continue working in the pole model together with current algebra
to compute both CF and SCS two-body weak decays of
$\Xi_c$ baryons.

In short, this work is motivated mainly by three parts:
(i) new data on the branching fractions and lifetimes of $\Xi_c^{+,0}$, (ii) correct sign predictions of $\alpha$ in $\Lambda_c^+\to \Sigma^+\pi^0$ and $\Sigma^0\pi^+$ by current algebra and (iii) the long-standing puzzle of $\Lambda_c^+\to \Xi^0K^+$ and its implication to the $\Xi_c$ sector.

This paper is organized as follows.  In Sec. \ref{sec:Formalism} we will set up
the formalism for computing branching fractions and up-down decay asymmetries, including
contributions from both
factorizable and nonfactorizable terms. Numerical results are
presented in Sec. \ref{sec:num}.
A conclusion will be given in Sec. \ref{sec:con}.
In Appendix \ref{app:wf}, we write down the baryon wave functions to fix our convention and then examine their behavior under $U$-, $V$-, and $I$-spin in Appendix \ref{app:UVI}.
Appendix \ref{app:FF} is devoted to  the form factors for $\Lambda_{c}^+\to \mathcal{B}$ transitions evaluated in the MIT bag model.
The expressions of baryon matrix elements and axial-vector form factor calculated in the MIT bag model will be presented in Appendix \ref{app:me}.

\section{Formalism}
\label{sec:Formalism}

\subsection{Kinematics}\label{sec:Kin}

Without loss of generality, the amplitude for the two-body weak decay  $\mathcal{B}_i\to\mathcal{B}_f P$
can be parameterized as
\begin{equation}
M(\mathcal{B}_i \to \mathcal{B}_f P)= i\ubar_f (A-B\gamma_5) u_i,
\end{equation}
where $P$ denotes a pseudoscalar meson.
Based on the $S$- and $P$- wave amplitudes, $A$ and $B$,
the decay width and up-down spin asymmetry  are given by
\begin{align}
& \Gamma=\frac{p_c}{8\pi}
\left[ \frac{(m_i+m_f)^2-m_P^2}{m_i^2}|A|^2
+\frac{(m_i-m_f)^2-m_P^2}{m_i^2}|B|^2\right], \label{eq:Gamma}\\
& \alpha=\frac{2\kappa {\rm{Re}} (A^* B)}{|A|^2+\kappa^2|B|^2}, \label{eq:alpha}
\end{align}
with $\kappa=p_c/(E_f+m_f)=\sqrt{(E_f-m_f)/(E_f+m_f)}$ and $p_c$ is the three-momentum
in the rest frame of mother particle.

The $S$- and $P$- wave amplitudes of the two-body decay
generically receive both factorizable and nonfactorizable contributions
\begin{align}
& A=A^{\rm{fac}}+A^{\rm{nf}},\qquad
B=B^{\rm{fac}}+B^{\rm{nf}}.
\label{eq:amplitude}
\end{align}
We should keep in mind that the above  formal decomposition is process dependent,
not all the processes contain both contributions shown in Eq. (\ref{eq:amplitude}).
To identify the explicit components, one way is to resort to
 the topological diagram method.
In the topological diagram approach, the external $W$-emission
and internal $W$-emission from the external quark are usually classified as
factorization contributions, while the nonfactorizable contributions arise from
inner $W$-emission and $W$-exchange diagrams.
Contrary to  weak decays of $\Lambda_c^+$ decay modes proceeding only through factorizable contributions cannot be found in $\Xi_{c}^{+,0}$ decays.

\subsection{Factorizable contribution}\label{sec:fac}


The description of the factorizable contribution of the charmed baryon decay $\mathcal{B}_c\to\mathcal{B} P$
is based on the effective Hamiltonian approach.

\subsubsection{General expression of factorizable amplitudes}
The effective Hamiltonian for CF process is
\begin{equation}
\mathcal{H}_{\rm{eff}}=\frac{G_F}{\sqrt{2}}V_{cs}V_{ud}^*(c_1O_1+c_2O_2)+h.c.,
\end{equation}
where the four-quark operators are given by
\begin{equation}
O_1=(\sbar c)(\ubar d),\quad
O_2=(\ubar c)(\sbar d),
\end{equation}
with $ (\qbar_1 q_2)\equiv \qbar_1\gamma_\mu(1-\gamma_5) q_2\nonumber$.
The
Wilson coefficients to the leading order are given as $c_1=1.346$ and $c_2=-0.636$ at $\mu=1.25\,\rm{GeV}$ and
$\Lambda_{\rm{MS}}^{(4)}=325\,{\rm{MeV}}$ \cite{Buchalla:1995vs}.
Under naive factorization
the amplitude can be written down as
\begin{equation}
M=\la P\mathcal{B}|\mathcal{H}_{\rm{eff}}|\mathcal{B}_{c}\ra
=\left\{\begin{array}{ll}
\frac{G_F}{\sqrt{2}}V_{cs}V_{ud}^* a_{2} \la P|(\sbar d)|0\ra \la \mathcal{B}|(\ubar c)|\mathcal{B}_{c}\ra, &P=\overline{K}^0, \\
\\
\frac{G_F}{\sqrt{2}}V_{cs}V_{ud}^* a_{1} \la P|(\ubar d)|0\ra \la \mathcal{B}|(\sbar c)|\mathcal{B}_{c}\ra , &P =\pi^+.
\end{array}
\label{eq:CF}
\right.
\end{equation}
where $a_1=c_1+\frac{c_2}{N}$ and $a_2=c_2+\frac{c_1}{N}$.
In terms of the decay constants
\footnote{Here we follow the PDG convention
$\la 0| A_\mu(0)|P(\bm{q})\ra =if_P q_\mu$ for the decay constant. This differs from the sign convention used in \cite{Cheng:2018hwl}.
}
\begin{equation}
\la K (q)|\sbar\gamma_\mu(1-\gamma_5) d|0\ra = if_K q_\mu,\quad
\la \pi (q)|\ubar\gamma_\mu(1-\gamma_5) d|0\ra = if_\pi q_\mu,
\label{KFF}
\end{equation}
and the form factors defined by
\begin{eqnarray}
\la \mathcal{B}(p_2)|\cbar\gamma_\mu(1-\gamma_5) u|\mathcal{B}_{c}(p_1)\ra
&=&\ubar_2 \left[ f_1(q^2) \gamma_\mu -f_2(q^2)i\sigma_{\mu\nu}\frac{q^\nu}{M}+f_3(q^2)\frac{q_\mu}{M}\right.\\
&&\hspace{0.5cm} -\left.\left(g_1(q^2)\gamma_\mu-g_2 (q^2)i\sigma_{\mu\nu}\frac{q^\nu}{M}+g_3(q^2)
\frac{q_\mu}{M}
\right)\gamma_5
\right]u_1,  \nonumber
\end{eqnarray}
with
the momentum transfer  $q=p_1-p_2$,
we obtain the  amplitude
\begin{equation}
M(\mathcal{B}_{c}\to \mathcal{B} P)
=i\frac{G_F}{\sqrt{2}}a_{1,2} V_{ud}^*V_{cs} f_P \ubar_2(p_2)\left[(m_1-m_2)f_1(q^2)
+(m_1+m_2)g_1(q^2) \gamma_5\right]u_1(p_1),
\end{equation}
where contributions from the form factors $f_{3}$ and $g_{3}$ can be neglected.
\footnote{To see the possible corrections from the form factors $f_3$ and $g_3$ for kaon or $\eta$ production in the final state, we notice that $m_P^2/m_{\Lambda_c}^2=0.047$ for the kaon and 0.057 for the $\eta$. Since the form factor $f_3$ is much smaller than $f_1$ (see e.g. Table IV of \cite{Zhao:2018zcb}), while $g_3$ is of the same order as $g_1$, it follows that the form factor $f_3$ can be safely neglected in the factorizable amplitude, while $g_3$ could make $\sim5\%$ corrections for kaon or $\eta$ production. For simplicity,  we will drop all the contributions from $f_3$ and $g_3$.
}
The factorizable contributions to $A$ and $B$ terms finally read
\begin{align}
&A^{\rm{fac}}\big|_{\rm{CF}}=\frac{G_F}{\sqrt{2}}a_{1,2} V_{ud}^*V_{cs} f_P(m_{\mathcal{B}_{c}}-m_{\mathcal{B}}) f_1(q^2),  
\nonumber\\
&B^{\rm{fac}}\big|_{\rm{CF}}= -\frac{G_F}{\sqrt{2}}a_{1,2} V_{ud}^*V_{cs} f_P(m_{\mathcal{B}_{c}}+m_{\mathcal{B}})
g_1(q^2),
\end{align}
where the choice of $a_i$ can be referred to Eq. (\ref{eq:CF}).

Likewise, the $S$- and $P$- wave amplitudes for SCS processes are given by
\begin{align} \label{eq:factSCS}
&A^{\rm{fac}}\big|_{\rm{SCS}}=\frac{G_F}{\sqrt{2}}a_{1,2} 
V_{uq}^*V_{cq} f_P(m_{\mathcal{B}_{c}}-m_{\mathcal{B}}) f_1(q^2),
\nonumber\\
&B^{\rm{fac}}\big|_{\rm{SCS}}= -\frac{G_F}{\sqrt{2}}a_{1,2} 
V_{uq}^*V_{cq} f_P(m_{\mathcal{B}_{c}}+m_{\mathcal{B}})
g_1(q^2),
\end{align}
where the flavor of the down-type quark $q$, $d$ or $s$, depends on the process.
If  $P=\eta_8$, both flavors contribute, for example,
\be \label{eq:facamp}
A^{\rm fac}(\Lambda_c^+\to p\eta) &=& {G_F\over \sqrt{2}}\,a_2 \left(V_{cs}V_{us}f_{\eta}^s+{1\over\sqrt{2}}V_{cd}V_{ud}f_{\eta}^q\right) (m_{\Lambda_c}-m_{p})f_1^{\Lambda_c p}(m_\eta^2),  \non \\
B^{\rm fac}(\Lambda_c^+\to p\eta) &=& -{G_F\over \sqrt{2}}\,a_2 \left(V_{cs}V_{us}f_{\eta}^s+{1\over\sqrt{2}}V_{cd}V_{ud}f_{\eta}^q\right) (m_{\Lambda_c}+m_{p})g_1^{\Lambda_c p}(m_\eta^2),
\en
where the decay constants are defined by
\be
\la \eta|\bar q\gamma_\mu(1-\gamma_5) q|0\ra=i{1\over \sqrt{2}}f_{\eta}^q q_\mu, \qquad
\la \eta|\bar s\gamma_\mu(1-\gamma_5) s|0\ra=if_{\eta}^s q_\mu.
\en
We shall follow \cite{Kroll} to use
$f_\eta^q= 107$ MeV and $f^s_\eta= -112$  MeV. Notice that in the case of $\pi^0$ production in the final state, one should replace $a_2$ by $-a_2/\sqrt{2}$ in the factorizable amplitude, where the extra factor of $-1/\sqrt{2}$ comes from the wave function of the $\pi^0$, $\pi^0=(u\bar u-d\bar d)/\sqrt{2}$.

\subsubsection{The parameterization of form factors}

There are two different non-perturbative parameters in factorizable amplitudes, the decay constant
and the form factor (FF).
There exist some efforts for estimating the FFs for $\Xi_c\to \mathcal{B}$ transition \cite{PerezMarcial:1989yh, Cheng:1991sn, Zhao:2018zcb,Faustov:2019ddj}. In this work we prefer to work out FFs for $\Xi_c$--${\cal B}$ transition and baryonic matrix elements all within the MIT bag model \cite{MIT}.
\footnote{See Chapter 18 of \cite{Close:1979} for a nice introduction to the MIT bag model. For the evaluation of baryon matrix elements and form factors in this model, see e.g. \cite{Cheng:1991sn,Cheng:1993gf}.
}
Since the decay rates and decay asymmetries are sensitive to the relative sign between factorizable and non-factorizable amplitudes, it is also desired to have an estimation of FFs in a globally consistent convention.

In this work we follow \cite{Fakirov:1977ta} to write the $q^2$ dependence of FF as
\begin{equation}
f_i(q^2)=\frac{f_i(0)}{(1-q^2/m_V^2)^2},\qquad
g_i(q^2)=\frac{g_i(0)}{(1-q^2/m_A^2)^2},
\label{eq:FF1}
\end{equation}
where $m_V
=2.01\,{\rm GeV}$, $m_A
=2.42\,{\rm GeV}$ for the  $(c\bar{d})$ quark content, and
$m_V=2.11\,{\rm GeV}$, $m_A=2.54\,{\rm GeV}$ for $(c\bar{s})$ quark content.
In the zero recoil limit where $q^2_{\rm{max}}=(m_i-m_f)^2$, FFs can be expressed within the  MIT bag model as \cite{Cheng:1993gf}
\begin{align} \label{eq:f1g1}
& f_1^{\mathcal{B}_f \mathcal{B}_i}(q^2_{\rm{max}})=\la \mathcal{B}_f^\uparrow| b_{q_1}^\dagger b_{q_2}| \mathcal{B}_i^\uparrow\ra \int d^3 \bm{r}\Big(u_{q_1}(r)u_{q_2}(r)+v_{q_1}(r)v_{q_2}(r)\Big),
\nonumber\\
& g_1^{\mathcal{B}_f \mathcal{B}_i}(q^2_{\rm{max}})=\la \mathcal{B}_f^\uparrow| b_{q_1}^\dagger b_{q_2} \sigma_z| \mathcal{B}_i^\uparrow\ra \int d^3 \bm{r}
\left(u_{q_1}(r)u_{q_2}(r)-\frac13v_{q_1}(r)v_{q_2}(r)\right),
\end{align}
where $u(r)$ and $v(r)$ are the large and small components, respectively, of the quark wave function in the bag model.
FFs at different $q^2$ are related via
\begin{equation}
f_i(q_2^2)=\frac{(1-q_1^2/m_V^2)^2}{(1-q_2^2/m_V^2)^2}f_i(q_1^2),\qquad
g_i(q_2^2)=\frac{(1-q_1^2/m_A^2)^2}{(1-q_2^2/m_A^2)^2}g_i(q_1^2).
\label{eq:FF2}
\end{equation}
This allows us to obtain the physical FF  at $q^2=m_P^2$.

It is obvious that the FF at $q^2_{\rm{max}}$ is determined only by the baryons in initial and final states.
However, its evolution with $q^2$ is governed by the relevant quark content. Such a dependence is reflected in
Table \ref{tab:FF}, in which the quark contents are shown in the second column.
In the zero recoil limit, the FFs at $q^2_{\rm{max}}$ calculated from Eq. (\ref{eq:FF1})
are presented in the third and sixth columns. And then in the fourth and seventh columns, the  evolution of FFs from $q^2=q^2_{\rm{max}}$ to $q^2=m_P^2$ are derived according to Eq. (\ref{eq:FF2}).
The bag integrals  $Y_{1,2}^{(s)}$ are defined by
\begin{align}
&Y_1=4\pi \int r^2 dr (u_u u_c+v_u v_c),\quad~~ Y_1^s=4\pi \int  r^2 dr (u_s u_c+v_s v_c), \nonumber\\
&Y_2=4\pi \int r^2 dr (u_u u_c-\frac13 v_u v_c),\quad Y_2^s=4\pi \int  r^2  dr (u_s u_c-\frac13 v_s v_c).
\end{align}
The model parameters are adopted from \cite{Cheng:2018hwl} and references therein. Numerically, we have $Y_1=0.88, Y_1^s=0.95, Y_2=0.77, Y_2^s=0.86$, which are consistent with
the corresponding numbers in \cite{Cheng:1993gf}.

\begin{table}[t!]
\linespread{1.5}
\footnotesize{
 \caption{Form factors of $\Xi_{c}^{0,+}\to \mathcal{B}P$ evaluated in the MIT bag model. The calculated results at $q^2=q^2_{\rm{max}}$ are presented in the third/sixth column. With different
 involved quark content shown in the second column, the evolution coefficients are shown in fourth/seventh column.
 The physical FFs $f_1(m_P^2)$ are shown  in the fifth column, likewise for $g_1(m_P^2)$.
  \label{tab:FF}}
\begin{ruledtabular}
\begin{tabular}{lcrcr |r c r}
 modes & ($c\bar{q}$)&  $f_1(q_{\rm{max}}^2)$ & $ f_1(m_P^2)/f_1(q_{\rm{max}}^2)$ &  $f_1(m_P^2)$ & $g_1(q_{\rm{max}}^2)$ &
 $g_1(m_P^2)/g_1(q_{\rm{max}}^2)$ & $g_1(m_P^2)$  \\
   \colrule
$\Xi_{c}^{+}\to\Sigma^{+} \overline{K}^0$ & $(c\bar{s})$  & $-\frac{\sqrt{6}}{2}Y_1$  &$0.44907$ & $-0.485$   & $-\frac{\sqrt{6}}{2}Y_2$  &   $0.60286$ & $-0.567$  \\
$\Xi_c^+\to\Xi^0\pi^+$  & $(c\bar{s})$  & $-\frac{\sqrt{6}}{2}Y_1^s$  &$0.49628$  & $-0.577$  & $-\frac{\sqrt{6}}{2}Y_2^s$  &$0.63416$  & $-0.667$ \\
\hline
$\Xi_c^0\to\Lambda\overline{K}^0$ & $(c\bar{s})$  & $\frac{1}{2}Y_1$  &   $0.38700$ & 0.171 & $\frac{1}{2}Y_2$  &$0.55337$  & 0.213 \\
$\Xi_c^0\to\Sigma^0\overline{K}^0$ & $(c\bar{s})$ & $\frac{\sqrt{3}}{2}Y_1$  &  $0.44929$  & 0.343 & $\frac{\sqrt{3}}{2}Y_2$  &
$0.60304$ & 0.401 \\
$\Xi_c^0\to\Xi^-\pi^+$  & $(c\bar{s})$ & $-\frac{\sqrt{6}}{2}Y_1^s$  & $0.49911$   & $-0.581$ & $-\frac{\sqrt{6}}{2}Y_2^s$  &$0.63636$ & $-0.669$ \\
\hline
$\Xi_c^+\to\Sigma^0\pi^+$ & $(c\bar{d})$ & $\frac{\sqrt{3}}{2}Y_1$  & $0.36045$   & 0.275 & $\frac{\sqrt{3}}{2}Y_2$  &$0.52523$ & 0.350 \\
$\Xi_c^+\to\Lambda\pi^+$ & $(c\bar{d})$ & $-\frac{1}{2}Y_1$  & $0.30260$  & $-0.134$  & $-\frac{1}{2}Y_2$  &$0.47622$ & $-0.183$ \\
$\Xi_c^+\to\Sigma^+\pi^0$ & $(c\bar{d})$ & $-\frac{\sqrt{6}}{2}Y_1$  &$0.35774$    & $-0.387$ & $-\frac{\sqrt{6}}{2}Y_2$  &$0.52294$ & $-0.492$ \\
$\Xi_c^+\to\Sigma^+\eta$ & $(c\bar{d})$& $-\frac{\sqrt{6}}{2}Y_1$  &  $0.41371$  & $-0.447$ & $-\frac{\sqrt{6}}{2}Y_2$
&$0.57735$ & $-0.543$ \\
$\Xi_c^+\to\Xi^0 K^+$ & $(c\bar{s})$& $-\frac{\sqrt{6}}{2}Y_1^s$  &$0.55058$    & $-0.641$ & $-\frac{\sqrt{6}}{2}Y_2^s$
 &$0.68080$ & $-0.716$ \\
\hline
$\Xi_c^0\to\Lambda\eta$ &  $(c\bar{d})$ & $\frac{1}{2}Y_1$  &  $0.34715$ & 0.153 & $\frac{1}{2}Y_2$
&$0.52343$ & 0.201 \\
$\Xi_c^0\to\Sigma^0\eta$&  $(c\bar{d})$ & $\frac{\sqrt{3}}{2}Y_1$  & $0.41395$ & 0.316
 & $\frac{\sqrt{3}}{2}Y_2$  &$0.57754$ & 0.384 \\
$\Xi_c^0\to\Lambda\pi^0$ & $(c\bar{d})$ & $\frac{1}{2}Y_1$  & $0.30019$ & 0.132  & $\frac{1}{2}Y_2$  &$0.47410$ & 0.182 \\
$\Xi_c^0\to\Sigma^0\pi^0$&  $(c\bar{d})$ & $\frac{\sqrt{3}}{2}Y_1$  & $0.35795$   & 0.274 & $\frac{\sqrt{3}}{2}Y_2$  &$0.52311$ & 0.348 \\
$\Xi_c^0\to\Sigma^-\pi^+$ &  $(c\bar{d})$ & $\frac{\sqrt{6}}{2}Y_1$  &$0.36183$    & 0.391 & $\frac{\sqrt{6}}{2}Y_2$  &$0.52638$ & 0.496 \\
$\Xi_c^0\to\Xi^- K^+$ & $ (c\bar{s})$  & $-\frac{\sqrt{6}}{2}Y_1^s$  & $0.55371$   & $-0.644$ & $-\frac{\sqrt{6}}{2}Y_2^s$  &$0.68316$ & $-0.719$ \\
\end{tabular}
\end{ruledtabular}}
\end{table}


\subsection{Nonfactorizable contribution}\label{sec:nf}
We work in the framework of the pole model to estimate nonfactorizable contributions.
It is known that the S-wave amplitude is dominated by the low-lying $1/2^-$ resonances, while the $P$-wave one governed by the ground-state $1/2^+$ pole.
The general formulas for $A$ ($S$-wave) and $B$ ($P$-wave) terms in the pole model are given by \cite{Cheng:1991sn}
\footnote{
Note that we have corrected the sign of the $B$ term in \cite{Cheng:2018hwl}.
}
\begin{align}
& A^{\rm{pole}}=-\sum\limits_{\mathcal{B}_n^*(1/2^-)}\left[\frac{g_{\mathcal{B}_f \mathcal{B}_n^* P}b_{n^* i}}{m_i-m_{n^*}} +
\frac{b_{fn^*}g_{\mathcal{B}_{n}^*\mathcal{B}_i P}}{m_f-m_{n^*}}\right],  \nonumber\\
& B^{\rm{pole}}=\sum\limits_{\mathcal{B}_n}\left[ \frac{g_{\mathcal{B}_f \mathcal{B}_n P}a_{ni}}{m_i-m_n}
+\frac{a_{fn}g_{\mathcal{B}_n \mathcal{B}_i P}}{m_f-m_n}\right],
\label{eq:pole}
\end{align}
where $a_{ij}, b_{ij}$ are the baryon matrix elements defined by
\begin{equation}
\la \mathcal{B}_n|\mathcal{H}|\mathcal{B}_i\ra = \ubar_n (a_{ni} - b_{n i}\gamma_5) u_i,\quad
\la \mathcal{B}_i^*(1/2^-) | \mathcal{H}|\mathcal{B}_j\ra =\ubar_{i^*} b_{i^* j} u_j.
\label{eq:amp}
\end{equation}
In the soft-meson limit, the intermediate excited $1/2^-$ states in the $S$-
wave amplitude can be summed up and reduced to a
commutator term \cite{Cheng:2018hwl},
\footnote{
The applied relation $ [Q_5, H^{\rm{pv}}_{\rm{eff}}]=-[Q, H^{\rm{pc}}_{\rm{eff}}]$ differs from that in \cite{Cheng:2018hwl} in sign.}
\begin{equation}
A^{\rm{com}}=-\frac{\sqrt{2}}{f_{P^a}}\la \mathcal{B}_f|[Q_5^a, H_{\rm{eff}}^{\rm{PV}}]|\mathcal{B}_i\ra
=\frac{\sqrt{2}}{f_{P^a}}\la \mathcal{B}_f|[Q^a, H_{\rm{eff}}^{\rm{PC}}]|\mathcal{B}_i\ra\label{eq:Apole}
\end{equation}
with
\begin{equation}
Q^a=\int d^3x \qbar\gamma^0\frac{\lambda^a}{2}q,\qquad
Q^a_5=\int d^3x \qbar\gamma^0\gamma_5\frac{\lambda^a}{2}q.
\end{equation}
By applying the generalized Goldberger-Treiman relation
\begin{equation} \label{eq:GT}
g_{_{\mathcal{B'B}P^a}}=\frac{\sqrt{2}}{f_{P^a}}(m_{\mathcal{B}}+m_{\mathcal{B'}})g^A_{\mathcal{B'B}},
\end{equation}
the $P$-wave amplitude can be simplified to
\begin{equation}
B^{\rm{pole}}=\frac{\sqrt{2}}{f_{P^a}}\sum_{\mathcal{B}_n}\left[ g^A_{\mathcal{B}_f \mathcal{B}_n}\frac{m_f+m_n}{m_i-m_n}a_{ni}
+a_{fn}\frac{m_i + m_n}{m_f-m_n} g_{\mathcal{B}_n \mathcal{B}_i}^A\right].
\label{eq:Bpole}
\end{equation}
Therefore, the two master equations Eq. (\ref{eq:Apole}) and Eq. (\ref{eq:Bpole}) for  the
nonfactorizable contributions in the pole model rely on
the commutator terms and the axial-vector form factor $g^A_{\mathcal{B'B}}$
which will be calculated in the MIT bag model in this work.


\subsubsection{$S$-wave amplitude}
\label{subsec:Swave}

We have deduced that the $S$-wave amplitude is determined by the commutator terms of conserving charge
$Q^a$ and the parity-conserving part of the Hamiltonian.
In the following
we list the expressions of $A^{\rm{com}}$
according to Eq. (\ref{eq:Apole}):
\begin{align}  \label{eq:commu}
&A^{\rm{com}}(\mathcal{B}_i\to \mathcal{B}_f \pi^{\pm})=\frac{1}{f_\pi}\la\mathcal{B}_f|[I_{\mp}, H_{\rm{eff}}^{\rm{PC}}]|\mathcal{B}_i\ra, \nonumber\\
&A^{\rm{com}}(\mathcal{B}_i\to \mathcal{B}_f \pi^{0})=\frac{\sqrt{2}}{f_\pi}\la \mathcal{B}_f|[I_3, H_{\rm{eff}}^{\rm{PC}}]|\mathcal{B}_i\ra,  \nonumber\\
&A^{\rm{com}}(\mathcal{B}_i\to \mathcal{B}_f \eta_8)=\sqrt{\frac32}\frac{1}{f_{\eta_8}}\la \mathcal{B}_f|[Y, H_{\rm{eff}}^{\rm{PC}}]|\mathcal{B}_i\ra,  \nonumber\\
&A^{\rm{com}}(\mathcal{B}_i\to \mathcal{B}_f K^{\pm})=\frac{1}{f_K}\la \mathcal{B}_f|[V_{\mp}, H_{\rm{eff}}^{\rm{PC}}]|\mathcal{B}_i\ra,  \\
&A^{\rm{com}}(\mathcal{B}_i\to \mathcal{B}_f \overline{K}^0 )=\frac{1}{f_K}\la \mathcal{B}_f|[U_{+}, H_{\rm{eff}}^{\rm{PC}}]|\mathcal{B}_i\ra,  \nonumber\\
&A^{\rm{com}}(\mathcal{B}_i\to \mathcal{B}_f {K^0})=\frac{1}{f_K}\la \mathcal{B}_f|[U_{-}, H_{\rm{eff}}^{\rm{PC}}]|\mathcal{B}_i\ra, \nonumber
\end{align}
where we have introduced the isospin, $U$-spin and $V$-spin ladder operators with
\be \label{eq:IUV}
I_+|d\ra=|u\ra, \quad I_-|u\ra=|d\ra, \quad U_+|s\ra=|d\ra, \quad U_-|d\ra=|s\ra, \quad V_+|s\ra=|u\ra, \quad V_-|u\ra=|s\ra. \nonumber \\
\en
In Eq. (\ref{eq:commu}),  $\eta_8$ is the octet component of the $\eta$ and $\eta'$
\be
\eta=\cos\theta\eta_8-\sin\theta\eta_0, \qquad \eta'=\sin\theta\eta_8+\cos\theta\eta_0,
\en
with $\theta=-15.4^\circ$ \cite{Kroll}. For the decay constant $f_{\eta_8}$,  we shall follow \cite{Kroll} to use $f_{\eta_8}=f_8\cos\theta$ with $f_8=1.26 f_\pi$.
Hypercharge $Y$, the conserving charge for processes involving $\eta_8$ in the final state, is taken to be  $Y=B+S-C$ as shown in \cite{Cheng:2018hwl}.
The baryon matrix elements of commutators in Eq. (\ref{eq:commu}), after the action of the ladder operators on
baryon wave functions shown in  Appendix \ref{app:UVI},  can be further reduced to pure matrix elements of
effective Hamiltonian, denoted by $a_{\mathcal{B}'\mathcal{B}}\equiv\la \mathcal{B}'|H_{\rm{eff}}^{\rm{PC}}|\mathcal{B}\ra$.
Then in terms of $a_{\mathcal{B}'\mathcal{B}}$,   nonfactorizable
contributions to $S$-wave amplitudes for charmed baryon decays are calculable.

For the  Cabibbo-favored processes, we have
\begin{align}
&A^{\rm{com}}(\Lambda_c^+\to p \overline{K}^0) =  -\frac{1}{f_{K}}a_{\Sigma^{+} \Lambda_{c}^{+}}, \qquad
A^{\rm{com}}(\Lambda_c^+\to \Lambda \pi^+) = 0,  \nonumber\\
&A^{\rm{com}}(\Lambda_c^+\to \Sigma^0 \pi^+) =  -\frac{\sqrt{2}}{f_{\pi}}a_{\Sigma^{+} \Lambda_{c}^{+}},  \quad~
A^{\rm{com}}(\Lambda_c^+\to \Sigma^+ \pi^0) = \frac{\sqrt{2}}{f_{\pi}}a_{\Sigma^{+} \Lambda_{c}^{+}},  \\
&A^{\rm{com}}(\Lambda_c^+\to \Xi^0 K^+) =\frac{1}{f_{K}}a_{\Sigma^{+} \Lambda_{c}^{+}}, \qquad A^{\rm{com}}(\Lambda_c^+\to \Sigma^+ \eta_8)
=\frac{\sqrt{2}}{\sqrt{3}f_{\eta_8}}a_{\Sigma_c^+\Sigma^+},  \nonumber
\end{align}
and
\begin{align} \label{eq:CFAcom}
&A^{\rm{com}}(\Xi_c^+\to \Sigma^+ \overline{K}^0)= \frac{1}{f_K} a_{\Sigma^+ \Lambda_c^+},
 \qquad
A^{\rm{com}}(\Xi_c^+\to \Xi^0 \pi^+)=  -\frac{1}{f_\pi} a_{\Xi^0 \Xi_c^0}, \nonumber \\
&A^{\rm{com}}(\Xi_c^0\to \Lambda \overline{K}^0)= \frac{1}{f_K} \frac{\sqrt{6}}{2} a_{\Xi^0 \Xi_c^0},  \qquad
A^{\rm{com}}(\Xi_c^0\to \Sigma^0 \overline{K}^0)=-\frac{1}{f_K} \frac{\sqrt{2}}{2} a_{\Xi^0 \Xi_c^0},  \nonumber\\
&A^{\rm{com}}(\Xi_c^0\to \Sigma^+ K^-)=\frac{1}{f_K} a_{\Xi^0 \Xi_c^0},  \qquad~~
A^{\rm{com}}(\Xi_c^0\to \Xi^0 \pi^0)= \frac{\sqrt{2}}{f_\pi} a_{\Xi^0 \Xi_c^0},    \\
&A^{\rm{com}}(\Xi_c^0\to \Xi^0 \eta_8)= \frac{\sqrt{6}}{f_{\eta_8}} a_{\Xi^0 \Xi_c^0},    \qquad \quad~~
A^{\rm{com}}(\Xi_c^0\to \Xi^- \pi^+)=    \frac{1}{f_\pi} a_{\Xi^0 \Xi_c^0}. \nonumber
\end{align}
For singly Cabibbo-suppressed processes we have
\begin{align} \label{eq:SCSAcom}
&A^{\rm{com}}(\Xi_c^+\to \Sigma^0 \pi^+)= -\frac{1}{f_\pi} \left( \sqrt{2} a_{\Sigma^+ \Xi_c^+}+a_{\Sigma^0 \Xi_c^0}\right),   \quad
A^{\rm{com}}(\Xi_c^+\to \Sigma^+ \pi^0)=  \frac{1}{\sqrt{2} f_\pi} a_{\Sigma^+ \Xi_c^+}, \nonumber \\
&A^{\rm{com}}(\Xi_c^+\to p \overline{K}^0)=  -\frac{1}{f_K}\left( a_{\Sigma^+ \Xi_c^+}-a_{p\Lambda_c^+}\right),   \qquad~~~
A^{\rm{com}}(\Xi_c^+\to \Lambda \pi^+)=-\frac{1}{f_\pi} a_{\Lambda \Xi_c^0},      \\
&A^{\rm{com}}(\Xi_c^+\to \Sigma^+ \eta_8)=  \frac{\sqrt{6}}{2}\frac{1}{f_{\eta_8}} a_{\Sigma^+ \Xi_c^+},      \qquad\qquad\qquad~~
A^{\rm{com}}(\Xi_c^+\to \Xi^0 K^+)=  \frac{1}{f_K} a_{\Sigma^+ \Xi_c^+}, \nonumber
\end{align}
and 
\begin{align} \label{eq:SCS0Acom}
&A^{\rm{com}}(\Xi_c^0\to \Xi^0 K^0)= \frac{1}{f_K} \left(
-\frac{\sqrt{2}}{2}a_{\Sigma^0 \Xi_c^0}+\frac{\sqrt{6}}{2} a_{\Lambda \Xi_c^0}   \right), \qquad~
A^{\rm{com}}(\Xi_c^0\to \Lambda \pi^0)= \frac{1}{\sqrt{2} f_{\pi}}a_{\Lambda \Xi_c^0}, \nonumber\\
&A^{\rm{com}}(\Xi_c^0\to \Xi^- K^+)=  -\frac{1}{f_K} \left(\frac{\sqrt{2}}{2}a_{\Sigma^0\Xi_c^0}
+\frac{\sqrt{6}}{2}a_{\Lambda \Xi_c^0}\right),   \qquad
A^{\rm{com}}(\Xi_c^0\to \Sigma^- \pi^+)= \frac{\sqrt{2}}{f_\pi} a_{\Sigma^0 \Xi_c^0}, \nonumber \\
&A^{\rm{com}}(\Xi_c^0\to p K^-)= -\frac{1}{f_K}
\left(\frac{\sqrt{2}}{2} a_{\Sigma^0 \Xi_c^0} +\frac{\sqrt{6}}{2} a_{\Lambda \Xi_c^0}
\right), \qquad\quad
A^{\rm{com}}(\Xi_c^0\to \Sigma^0 \pi^0)= \frac{1}{\sqrt{2} f_{\pi}}a_{\Sigma^0 \Xi_c^0}, \nonumber\\
&A^{\rm{com}}(\Xi_c^0\to n \overline{K}^0)=   \frac{1}{f_K} \left(\frac{\sqrt{2}}{2} a_{\Sigma^0 \Xi_c^0}
-\frac{\sqrt{6}}{2} a_{\Lambda \Xi_c^0}\right),   \qquad\qquad
A^{\rm{com}}(\Xi_c^0\to \Sigma^+ \pi^-)= - \frac{\sqrt{2}}{f_\pi} a_{\Sigma^0 \Xi_c^0}. \nonumber \\
&A^{\rm{com}}(\Xi_c^0\to \Lambda \eta_8)= \frac{\sqrt{6}}{2}\frac{1}{f_{\eta_8}} a_{\Lambda \Xi_c^0},  \qquad\qquad\qquad\qquad\qquad\quad~
A^{\rm{com}}(\Xi_c^0\to \Sigma^0 \eta_8)= \frac{\sqrt{6}}{2}\frac{1}{f_{\eta_8}} a_{\Sigma^0 \Xi_c^0}.
\end{align}
The nonfactorizable $S$-wave amplitudes for SCS decays of $\Lambda_c^+$ can be found in \cite{Cheng:2018hwl}.
The evaluation of the baryon matrix elements $a_{\mathcal{B'B}}$ in the MIT bag model and results are presented in Appendix \ref{HME}.

\subsubsection{$P$-wave amplitude}
Through the generalized Goldberger-Treiman relation Eq. (\ref{eq:GT}),
the strong coupling of $\mathcal{B'B}M$ can be expressed in terms of the axial-vector form factor $g^A_{\mathcal{B'B}}$.
Based on Eq.(\ref{eq:Bpole}), $P$-wave amplitudes are given
as follows. For Cabibbo-favored processes we have
\be
B^{\rm{ca}}(\Lambda_c^+\to p \overline{K}^0) &=& \frac{1}{f_{K}}\left(g^{(\overline{K}^{0})}_{p\Sigma^{+}}\frac{m_{p}+m_{\Sigma^{+}}}
{m_{\Lambda^{+}_{c}}-m_{\Sigma^{+}}}a_{\Sigma^{+}\Lambda^{+}_{c}}\right), \nonumber\\
B^{\rm{ca}}(\Lambda_c^+\to \Lambda \pi^+) &=&  \frac{1}{f_{\pi}}\left(a_{\Lambda \Sigma^{0}_{c}}\frac{m_{\Lambda_{c}^{+}}+m_{\Sigma^{0}_{c}}}
{m_{\Lambda}-m_{\Sigma^{0}_{c}}}g^{A(\pi^{+})}_{\Sigma_{c}^{0}\Lambda^{+}_{c}}+g^{A(\pi^{+})}_{\Lambda \Sigma^{+}}\frac{m_{\Lambda}+m_{\Sigma^{+}}}
{m_{\Lambda_{c}^{+}}-m_{\Sigma^{+}}}a_{ \Sigma^{+}\Lambda_{c}^{+}}\right), \nonumber\\
B^{\rm{ca}}(\Lambda_c^+\to \Sigma^0 \pi^+) &=&   \frac{1}{f_{\pi}}\left(a_{\Sigma^{0} \Sigma^{0}_{c}}\frac{m_{\Lambda_{c}^{+}}+m_{\Sigma^{0}_{c}}}
{m_{\Sigma^{0}}-m_{\Sigma^{0}_{c}}}g^{A(\pi^{+})}_{\Sigma_{c}^{0}\Lambda^{+}_{c}}+g^{A(\pi^{+})}_{\Sigma^{0} \Sigma^{+}}\frac{m_{\Sigma^{0}}+m_{\Sigma^{+}}}
{m_{\Lambda_{c}^{+}}-m_{\Sigma^{+}}}a_{ \Sigma^{+}\Lambda_{c}^{+}}\right),  \nonumber\\
B^{\rm{ca}}(\Lambda_c^+\to \Sigma^+ \pi^0) &=&  \frac{\sqrt{2}}{f_{\pi}}\Bigg(a_{\Sigma^{+} \Lambda^{+}_{c}}\frac{m_{\Lambda_{c}^{+}}+m_{\Lambda_{c}^{+}}}
{m_{\Sigma^{+}}-m_{\Lambda_{c}^{+}}}g^{A(\pi^{0})}_{\Lambda_{c}^{+}\Lambda^{+}_{c}}+a_{\Sigma^{+} \Sigma^{+}_{c}}\frac{m_{\Lambda_{c}^{+}}+m_{\Sigma_{c}^{+}}}
{m_{\Sigma^{+}}-m_{\Sigma_{c}^{+}}}g^{A(\pi^{0})}_{\Sigma_{c}^{+}\Lambda^{+}_{c}} \nonumber\\
&+& g^{A(\pi^{0})}_{\Sigma^{+}\Sigma^{+}}\frac{m_{\Sigma^{+}}+m_{\Sigma^{+}}}
{m_{\Lambda_{c}^{+}}-m_{\Sigma^{+}}}a_{ \Sigma^{+}\Lambda_{c}^{+}}\Bigg),  \\
B^{\rm{ca}}(\Lambda_c^+\to \Xi^0 K^+) &=& \frac{1}{f_{K}}\left(g^{A({K}^{+})}_{\Xi^{0}\Sigma^{+}}\frac{m_{\Xi^{0}}+m_{\Sigma^{+}}}
{m_{\Lambda^{+}_{c}}-m_{\Sigma^{+}}}a_{\Sigma^{+}\Lambda^{+}_{c}}\right), \nonumber
\en
and
\be \label{eq:CFBca}
B^{\rm{ca}}(\Xi_c^+\to \Sigma^+ \overline{K}^0) &=&
\frac{1}{f_K}\left( a_{\Sigma^+ \Lambda_c^+}\frac{m_{\Xi_c^+}+m_{\Lambda_c^+}}{m_{\Sigma^+}-m_{\Lambda_c^+}}
g^{A(\overline{K}^0)}_{\Lambda_c^+ \Xi_c^+}
+
a_{\Sigma^+ \Sigma_c^+}\frac{m_{\Xi_c^+}+m_{\Sigma_c^+}}{m_{\Sigma^+}-m_{\Sigma_c^+}}
g^{A(\overline{K}^0)}_{\Sigma_c^+ \Xi_c^+}
\right),
 \nonumber\\
B^{\rm{ca}}(\Xi_c^+\to \Xi^0 \pi^+) & = &\frac{1}{f_\pi}
\left( a_{\Xi^0 \Xi_c^0}\frac{m_{\Xi_c^+}+m_{\Xi_c^0}}{m_{\Xi^0}-m_{\Xi_c^0}} g^{A(\pi^+)}_{\Xi_c^0
\Xi_c^+}+
a_{\Xi^0 \Xi_c^{'0}}\frac{m_{\Xi_c^+}+m_{\Xi_c^{'0}}}{m_{\Xi^0}-m_{\Xi_c^{'0}}} g^{A(\pi^+)}_{\Xi_c^{'0}
\Xi_c^+}
  \right), \nonumber\\
B^{\rm{ca}}(\Xi_c^0\to \Lambda \overline{K}^0)&=& \frac{1}{f_K} \left(
a_{\Lambda \Sigma_c^0} \frac{m_{\Xi_c^0}+m_{\Sigma_c^0}}{m_{\Lambda}-m_{\Sigma_c^0}}
g^{A(\overline{K}^0)}_{\Sigma_c^0 \Xi_c^0}
+g^{A(\overline{K}^0)}_{\Lambda \Xi^0} \frac{m_\Lambda + m_{\Xi^0}}{m_{\Xi_c^0}-m_{\Xi^0}} a_{\Xi^0 \Xi_c^0}\right),
  \nonumber\\
B^{\rm{ca}}(\Xi_c^0\to \Sigma^0 \overline{K}^0) & =&     \frac{1}{f_K} \left(
a_{\Sigma^0 \Sigma_c^0} \frac{m_{\Xi_c^0}+m_{\Sigma_c^0}}{m_{\Sigma^0}-m_{\Sigma_c^0}}
g^{A(\overline{K}^0)}_{\Sigma_c^0 \Xi_c^0}
+g^{A(\overline{K}^0)}_{\Sigma^0 \Xi^0} \frac{m_{\Sigma^0} + m_{\Xi^0}}{m_{\Xi_c^0}-m_{\Xi^0}} a_{\Xi^0 \Xi_c^0}\right),
  \nonumber\\
B^{\rm{ca}}(\Xi_c^0\to \Sigma^+ K^-)&=& \frac{1}{f_K} \left( g^{A(K^-)}_{\Sigma^+ \Xi^0}
\frac{m_{\Sigma^+}+m_{\Xi^0}}{m_{\Xi_c^0}-m_{\Xi^0}} a_{\Xi^0 \Xi_c^0}
\right),    \\
B^{\rm{ca}}(\Xi_c^0\to \Xi^0 \pi^0) & =& \frac{\sqrt{2}}{f_\pi}\Bigg( a_{\Xi^0 \Xi_c^0}\frac{m_{\Xi_c^0}+m_{\Xi_c^0}}{m_{\Xi^0}
-m_{\Xi_c^0}}g^{A(\pi^0)}_{\Xi_c^0 \Xi_c^0}+
a_{\Xi^0 \Xi_c^{'0}}\frac{m_{\Xi_c^0}+m_{\Xi_c^{'0}}}{m_{\Xi^0}
-m_{\Xi_c^{'0}}}g^{A(\pi^0)}_{\Xi_c^{'0} \Xi_c^0}
\nonumber \\
&& +g^{A(\pi^0)}_{\Xi^0 \Xi^0}\frac{m_{\Xi^0}+m_{\Xi^0}}{m_{\Xi_c^0}-m_{\Xi^0}}a_{\Xi^0 \Xi_c^0} \Bigg),
    \nonumber\\
B^{\rm{ca}}(\Xi_c^0\to \Xi^0 \eta_8)&=&
\frac{\sqrt{2}}{f_{\eta_8}}\Bigg( a_{\Xi^0 \Xi_c^0}\frac{m_{\Xi_c^0}+m_{\Xi_c^0}}{m_{\Xi^0}
-m_{\Xi_c^0}}g^{A(\eta_8)}_{\Xi_c^0 \Xi_c^0}+
a_{\Xi^0 \Xi_c^{'0}}\frac{m_{\Xi_c^0}+m_{\Xi_c^{'0}}}{m_{\Xi^0}
-m_{\Xi_c^{'0}}}g^{A(\eta_8)}_{\Xi_c^{'0} \Xi_c^0} \nonumber \\
&& +g^{A(\eta_8)}_{\Xi^0 \Xi^0}\frac{m_{\Xi^0}+m_{\Xi^0}}{m_{\Xi_c^0}-m_{\Xi^0}}a_{\Xi^0 \Xi_c^0}
\Bigg), \nonumber \\
B^{\rm{ca}}(\Xi_c^0\to \Xi^- \pi^+) & =&    \frac{1}{f_\pi}\left(
g^{A(\pi^+)}_{\Xi^-\Xi^0}\frac{m_{\Xi^-}+m_{\Xi^0}}{m_{\Xi_c^0}-m_{\Xi^0}}a_{\Xi^0 \Xi_c^0}
\right). \nonumber
\en
The $P$-wave amplitudes for singly Cabibbo-suppressed processes read
\be \label{eq:SCSBca}
B^{\rm{ca}}(\Xi_c^+\to \Lambda \pi^+) &=& \frac{1}{f_\pi} \Bigg(
g^{A(\pi^+)}_{\Lambda \Sigma^+} \frac{m_{\Lambda}+m_{\Sigma^+}}{m_{\Xi_c^+}-m_{\Sigma^+}}a_{\Sigma^+ \Xi_c^+}
+
a_{\Lambda \Xi_c^0}\frac{m_{\Xi_c^+}+m_{\Xi_c^0}}{m_{\Lambda}-m_{\Xi_c^0}}g^{A(\pi^+)}_{\Xi_c^0 \Xi_c^+}
 \nonumber \\
&& +a_{\Lambda \Xi_c^{'0}}\frac{m_{\Xi_c^+}+m_{\Xi_c^{'0}}}{m_{\Lambda}-m_{\Xi_c^{'0}}}g^{A(\pi^+)}_{\Xi_c^{'0} \Xi_c^+}
\Bigg),      \nonumber\\
B^{\rm{ca}}(\Xi_c^+\to \Sigma^0 \pi^+) &=& \frac{1}{f_\pi} \Bigg(
g^{A(\pi^+)}_{\Sigma^0 \Sigma^+} \frac{m_{\Sigma^0}+m_{\Sigma^+}}{m_{\Xi_c^+}-m_{\Sigma^+}}a_{\Sigma^+ \Xi_c^+}
+
a_{\Sigma^0 \Xi_c^0}\frac{m_{\Xi_c^+}+m_{\Xi_c^0}}{m_{\Sigma^0}-m_{\Xi_c^0}}g^{A(\pi^+)}_{\Xi_c^0 \Xi_c^+}
\nonumber \\
&& +a_{\Sigma^0 \Xi_c^{'0}}\frac{m_{\Xi_c^+}+m_{\Xi_c^{'0}}}{m_{\Sigma^0}-m_{\Xi_c^{'0}}}g^{A(\pi^+)}_{\Xi_c^{'0} \Xi_c^+}
\Bigg),
  \nonumber\\
B^{\rm{ca}}(\Xi_c^+\to \Sigma^+ \pi^0) &=&  \frac{\sqrt{2}}{f_\pi} \left(
g^{A(\pi^0)}_{\Sigma^+ \Sigma^+} \frac{m_{\Sigma^+}+m_{\Sigma^+}}{m_{\Xi_c^+}-m_{\Sigma^+}}a_{\Sigma^+ \Xi_c^+}
\right), \\
B^{\rm{ca}}(\Xi_c^+\to \Sigma^+ \eta_8) &=&  \frac{\sqrt{2}}{f_{\eta_8}}
 \Bigg(
g^{A(\eta_8)}_{\Sigma^+ \Sigma^+} \frac{m_{\Sigma^+}+m_{\Sigma^+}}{m_{\Xi_c^+}-m_{\Sigma^+}}a_{\Sigma^+ \Xi_c^+}
+ a_{\Sigma^+ \Xi_c^+} \frac{m_{\Xi_c^+}+m_{\Xi_c^+ }}{m_{ \Sigma^+}-m_{\Xi_c^+}} g^{A(\eta_8)}_{\Xi_c^+ \Xi_c^+}
\nonumber \\
&& + a_{\Sigma^+ \Xi_c^{'+}} \frac{m_{\Xi_c^+}+m_{\Xi_c^{'+} }}{m_{ \Sigma^+}-m_{\Xi_c^{'+}}} g^{A(\eta_8)}_{\Xi_c^{'+} \Xi_c^+}
\Bigg),
  \nonumber\\
B^{\rm{ca}}(\Xi_c^+\to p \overline{K}^0) &=& \frac{1}{f_K} \Bigg(
g^{A(\overline{K}^0)}_{p\Sigma^+} \frac{m_p + m_{\Sigma^+}}{m_{\Xi_c^+}-m_{\Sigma^+}}a_{\Sigma^+ \Xi_c^+}
+ a_{p\Sigma_c^+} \frac{m_{\Xi_c^+}+m_{\Sigma_c^+}}{m_p -m_{\Sigma_c^+}} g^{A(\overline{K}^0)}_{\Sigma_c^+ \Xi_c^+} \nonumber \\
&& + a_{p\Lambda_c^+} \frac{m_{\Xi_c^+}+m_{\Lambda_c^+}}{m_p -m_{\Lambda_c^+}} g^{A(\overline{K}^0)}_{\Lambda_c^+ \Xi_c^+}
\Bigg), \nonumber\\
B^{\rm{ca}}(\Xi_c^+\to \Xi^0 K^+) &=&  \frac{1}{f_K} \left(g^{A(K^+)}_{\Xi^0 \Sigma^+}\frac{m_{\Xi^0}+m_{\Sigma^+}}{m_{\Xi_c^+}-m_{\Sigma^+}}a_{\Sigma^+ \Xi_c^+}+
a_{\Xi^0 \Omega_c^0}\frac{m_{\Xi_c^+}+m_{\Omega_c^0}}{m_{\Xi^0}-m_{\Omega_c^0}}
g^{A(K^+)}_{\Omega_c^0 \Xi_c^+}
\right),  \nonumber
\en
and
\be  \label{eq:SCS0Bca}
B^{\rm{ca}}(\Xi_c^0\to \Lambda \pi^0) &=&  \frac{\sqrt{2}}{f_\pi}\left(
g^{A(\pi^0)}_{\Lambda \Sigma^0}\frac{m_{\Lambda}+m_{\Sigma^0}}{m_{\Xi_c^0}
-m_{\Sigma^0}}a_{\Sigma^0 \Xi_c^0}
+
g^{A(\pi^0)}_{\Lambda \Lambda}\frac{{{m_{\Lambda}}}+m_{\Lambda}}{m_{\Xi_c^0}
-m_{\Lambda}}a_{\Lambda \Xi_c^0}
\right),   \nonumber\\
B^{\rm{ca}}(\Xi_c^0\to \Lambda \eta_8) &=&  \frac{\sqrt{2}}{f_{\eta_8}}\Bigg(
g^{A(\eta_8)}_{\Lambda \Sigma^0}\frac{m_{\Lambda}+m_{\Sigma^0}}{m_{\Xi_c^0}
-m_{\Sigma^0}}a_{\Sigma^0 \Xi_c^0}
+
g^{A(\eta_8)}_{\Lambda \Lambda}\frac{{{m_{\Lambda}}}+m_{\Lambda}}{m_{\Xi_c^0}
-m_{\Lambda}}a_{\Lambda \Xi_c^0} \nonumber \\
&+&  a_{\Lambda \Xi_c^0}\frac{m_{\Xi_c^0}+m_{\Xi_c^0}}{m_{\Lambda}-m_{\Xi_c^0}}
g^{A(\eta_8)}_{\Xi_c^0 \Xi_c^0} +a_{\Lambda {\Xi'}_c^0}\frac{m_{\Xi_c^0}+m_{{\Xi'}_c^0}}{m_{\Lambda}-m_{{\Xi'}_c^0}}
g^{A(\eta_8)}_{{\Xi'}_c^0 \Xi_c^0}\Bigg),  \nonumber \\
B^{\rm{ca}}(\Xi_c^0\to \Sigma^0 \pi^0) &=&  \frac{\sqrt{2}}{f_\pi}\left(
g^{A(\pi^0)}_{\Sigma^0 \Sigma^0}\frac{m_{\Sigma^0}+m_{\Sigma^0}}{m_{\Xi_c^0}
-m_{\Sigma^0}}a_{\Sigma^0 \Xi_c^0}
+
g^{A(\pi^0)}_{\Sigma^0 \Lambda}\frac{m_{\Sigma^0}+m_{\Lambda}}{m_{\Xi_c^0}
-m_{\Lambda}}a_{\Lambda \Xi_c^0}
\right),
\nonumber\\
B^{\rm{ca}}(\Xi_c^0\to \Sigma^0 \eta_8) &=&  \frac{\sqrt{2}}{f_{\eta_8}}\Bigg(
g^{A(\eta_8)}_{\Sigma^0 \Sigma^0}\frac{m_{\Sigma^0}+m_{\Sigma^0}}{m_{\Xi_c^0}
-m_{\Sigma^0}}a_{\Sigma^0 \Xi_c^0}
+
g^{A(\eta_8)}_{\Sigma^0 \Lambda}\frac{m_{\Sigma^0}+m_{\Lambda}}{m_{\Xi_c^0}
-m_{\Lambda}}a_{\Lambda \Xi_c^0}   \nonumber\\
&+&  a_{\Sigma^0 \Xi_c^0}\frac{m_{\Xi_c^0}+m_{\Xi_c^0}}{m_{\Sigma^0}-m_{\Xi_c^0}}
g^{A(\eta_8)}_{\Xi_c^0 \Xi_c^0} +a_{\Sigma^0 {\Xi'}_c^0}\frac{m_{\Xi_c^0}+m_{{\Xi'}_c^0}}{m_{\Sigma^0}-m_{{\Xi'}_c^0}}
g^{A(\eta_8)}_{{\Xi'}_c^0 \Xi_c^0}\Bigg),  \nonumber \\
B^{\rm{ca}}(\Xi_c^0\to \Sigma^- \pi^+) &=&
\frac{1}{f_\pi}\left(
g^{A(\pi^+)}_{\Sigma^- \Sigma^0}\frac{m_{\Sigma^-}+m_{\Sigma^0}}{m_{\Xi_c^0}
-m_{\Sigma^0}}a_{\Sigma^0 \Xi_c^0}
+
g^{A(\pi^+)}_{\Sigma^- \Lambda}\frac{m_{\Sigma^-}+m_{\Lambda}}{m_{\Xi_c^0}
-m_{\Lambda}}a_{\Lambda \Xi_c^0}
\right),    \\
B^{\rm{ca}}(\Xi_c^0\to \Sigma^+ \pi^-)  &=&  \frac{1}{f_\pi}\left(
g^{A(\pi^-)}_{\Sigma^+ \Sigma^0} \frac{m_{\Sigma^+}+m_{\Sigma^0}}{m_{\Xi_c^0}-m_{\Sigma^0}}
a_{\Sigma^0 \Xi_c^0}+
g^{A(\pi^-)}_{\Sigma^+ \Lambda} \frac{m_{\Sigma^+}+m_{\Lambda}}{m_{\Xi_c^0}-m_{\Lambda}}
a_{\Lambda \Xi_c^0}
\right).
  \nonumber \\
B^{\rm{ca}}(\Xi_c^0\to p K^-) &=& \frac{1}{f_K} \left( g^{A(K^-)}_{p\Sigma^0}
\frac{m_p + m_{\Sigma^0}}{m_{\Xi_c^0}-m_{\Sigma^0}}a_{\Sigma^0 \Xi_c^0}
+
g^{A(K^-)}_{p\Lambda}
\frac{m_p + m_{\Lambda}}{m_{\Xi_c^0}-m_{\Lambda}}a_{\Lambda \Xi_c^0}
\right),
 \nonumber\\
B^{\rm{ca}}(\Xi_c^0\to n \overline{K}^0) &=& \frac{1}{f_K} \left( g^{A(\overline{K}^0)}_{n \Sigma^0}
\frac{m_n + m_{\Sigma^0}}{m_{\Xi_c^0}-m_{\Sigma^0}}a_{\Sigma^0 \Xi_c^0}
+
g^{A(\overline{K}^0)}_{n\Lambda}
\frac{m_n + m_{\Lambda}}{m_{\Xi_c^0}-m_{\Lambda}}a_{\Lambda \Xi_c^0}
+
a_{n\Sigma_c^0}\frac{m_{\Xi_c^0}+m_{\Sigma_c^0}}{m_n-m_{\Sigma_c^0}}
g^{A(\overline{K}^0)}_{\Sigma_c^0 \Xi_c^0}
\right),   \nonumber\\
B^{\rm{ca}}(\Xi_c^0\to \Xi^0 K^0) &=& \frac{1}{f_K}
\left( a_{\Xi^0 \Omega_c^0}\frac{m_{\Xi_c^0}+m_{\Omega_c^0}}{m_{\Xi^0}-m_{\Omega_c^0}}
g^{A(K^0)}_{\Omega_c^0 \Xi_c^0}
+g^{A(K^0)}_{\Xi^0 \Sigma^0}\frac{m_{\Xi^0} + m_{\Sigma^0}}{m_{\Xi_c^0}-m_{\Sigma^0}}a_{\Sigma^0 \Xi_c^0}
+g^{A(K^0)}_{\Xi^0 \Lambda}\frac{m_{\Xi^0} + m_{\Lambda}}{m_{\Xi_c^0}-m_{\Lambda}}a_{\Lambda \Xi_c^0}
\right),
 \nonumber\\
B^{\rm{ca}}(\Xi_c^0\to \Xi^- K^+) &=& \frac{1}{f_K}
\left( g^{A(K^+)}_{\Xi^-\Sigma^0} \frac{m_{\Xi^-}+m_{\Sigma^0}}{m_{\Xi_c^0}-m_{\Sigma^0}}
a_{\Sigma^0 \Xi_c^0}
+
g^{A(K^+)}_{\Xi^-\Lambda} \frac{m_{\Xi^-}+m_{\Lambda}}{m_{\Xi_c^0}-m_{\Lambda}}
a_{\Lambda \Xi_c^0}
\right). \nonumber
\en
The nonfactorizable $P$-wave amplitudes for SCS decays of $\Lambda_c^+$ can be found in \cite{Cheng:2018hwl}.
In addition to the baryon matrix element $a_{\mathcal{BB'}}$, another quantity
in the nonfactorizable part of $P$-wave amplitude is the axial-vector form factor $g^{A(P)}_{\mathcal{B'B}}$.
For consistency, the estimation of $g^{A(P)}_{\mathcal{B'B}}$ is carried out in the MIT bag model and the results
are shown in Sec. \ref{sec:axial-vectorFF}.
As seen in the next section, one of the $W$-exchange diagrams, the so-called type-III diagram in which the quark pair is produced between the two
quark lines  without $W$-exchange, does not contribute to the nonfactorizable $S$- and $P$-wave amplitudes.
This will be discussed in detail there.

\section{Numerical results and discussions}
\label{sec:num}

\subsection{$\Lambda_c^+$ decays}

Before proceeding to the $\Xi_c$ sector, we first discuss $\Lambda_c^+$ decays as the measurements of branching fractions and decay asymmetries are well established for many of the channels. The goal is to see what we can learn from the $\Lambda_c^+$ physics. We show in Table \ref{tab:LambdacCF} the results of calculations for CF and SCS $\Lambda_c^+$ decays.
For the form factors $f_1$ and $g_1$,  we  follow \cite{Gutsche:2014zna} to use
\footnote{The sign of the form factors is fixed by Eq. (\ref{eq:f1g1}). }
\be
f_1^{\Lambda_c p}(0)=-0.470, \qquad g_1^{\Lambda_c p}(0)=-0.414\,,
\en
for $\Lambda_c$--$p$ transition and  rescale the form factors for $\Lambda_c$--$\Lambda$ transition to fit the decay  $\Lambda_c^+\to\Lambda \pi^+$ so that $f_1^{\Lambda_c\Lambda}(0)=0.406$ and $g_1^{\Lambda_c\Lambda}(0)=0.370$.
\footnote{We have checked if the form factors for $\Lambda_c^+$--$p$ and $\Lambda_c^+$--$\Lambda$ transitions given in Appendix \ref{app:FF} are used, the resulting decay asymmetries will remain stable, but the calculated branching fractions are not as good as those shown in Table \ref{tab:LambdacCF} but within a factor of 2. }
We see from Table \ref{tab:LambdacCF} that the calculated branching fractions and decay asymmetries are in general consistent with experiment except for the decay asymmetry in the decay $\Lambda_c^+\to p\overline{K}^0$. While all the predictions of $\alpha(\Lambda_c^+\to p\overline{K}^0)$ in the literature are all negative except \cite{Xu:1992vc}, the measured  asymmetry by BESIII turns out to be positive with a large uncertainty, $0.18\pm0.45$ \cite{Ablikim:2019zwe}. This issue needs to be resolved in future study.

\begin{table}[t]
\scriptsize{
\caption{ The predicted $S$- and $P$-wave amplitudes of Cabibbo-favored (upper entry) and singly Cabibbo-suppressed  (lower entry) $\Lambda_c^+\to{\cal B}+P$ decays in units of $10^{-2} G_F{\rm GeV}^2$. Branching fractions and the asymmetry parameter $\alpha$ are shown in the last four columns. Experimental results for decay asymmetries are taken from \cite{Ablikim:2019zwe} except the modes $\Lambda\pi^+$ and $\Sigma^+\pi^0$ where the world averages are obtained from \cite{Ablikim:2019zwe} and \cite{Tanabashi:2018oca}.
} \label{tab:LambdacCF}
\vspace{0.3cm}
\begin{center}
\begin{ruledtabular}
\begin{tabular}
{lrrrrrr |ccrc}
 Channel & $A^{\rm{fac}}$ &  $A^{\rm{com}}$ & $A^{\rm{tot}}$ & $B^{\rm{fac}}$ &  $B^{\rm{ca}}$ & $B^{\rm{tot}}$ & $\mathcal{B}_{\rm{theo}}$
 & $\mathcal{B}_{\rm{exp}}$ \cite{Tanabashi:2018oca} &  $\alpha_{\rm{theo}}$&  $\alpha_{\rm{exp}}$\\
 \hline
 \colrule
$\Lambda_{c}^{+}\to p \overline{K}^{0}$ & $3.45$  & $4.48$  &$7.93$    & $-6.98$  &   $-2.06$   &$-9.04$     & $2.11\times 10^{-2}$  & $(3.18\pm0.16)10^{-2}$   & $-0.75$ & ~~$0.18\pm0.45$   \\
$\Lambda_{c}^{+}\to \Lambda \pi^{+}$ & $5.34$  & $0$  &$5.34$    & $-14.11$  &   $3.60$   &$-10.51$     & $1.30\times 10^{-2}$  & $(1.30\pm0.07)10^{-2}$     & $-0.93$ & $-0.84\pm0.09$    \\
$\Lambda_{c}^{+}\to \Sigma^{0} \pi^{+}$    & $0$  & $7.68$  &$7.68$    & $0$  &   $-11.38$   &$-11.38$  & $2.24\times 10^{-2}$ & $(1.29\pm0.07)10^{-2}$  & $-0.76$ & $-0.73\pm0.18$    \\
$\Lambda_{c}^{+}\to \Sigma^{+} \pi^{0}$    & $0$   & $-7.68$  &$-7.68$    & $0$  &   $11.34$   &$11.34$  & $2.24\times 10^{-2}$ & $(1.25\pm0.10)10^{-2}$   & $-0.76$ & $-0.55\pm0.11$    \\
$\Lambda_{c}^{+}\to \Xi^{0} K^{+}$    & $0$  & $-4.48$  &$-4.48$    & $0$  &   $-12.10$   &$-12.10$  & $0.73\times 10^{-2}$  & $(0.55\pm0.07)10^{-2}$ & $0.90$ &      \\
$\Lambda_{c}^{+}\to \Sigma^{+} \eta$    & $0$  & $3.10$  &$3.10$    & $0$  &   $-15.54$   &$-15.54$  & $0.74\times 10^{-2}$  & $(0.53\pm0.15)10^{-2}$ & $-0.95$      \\
\hline
$\Lambda_c^+\to p\pi^0$ &  $0.41$ & $-0.81$ & $-0.40$ & $-0.87$ & $2.07$ & $1.21$ & $1.26\times 10^{-4}$ & $<2.7\times 10^{-4}$ & $-0.97$ \\
$\Lambda_c^+\to p\eta$ & $-0.96$ & $-1.11$ & $-2.08$ & $1.93$ & $-0.34$ & $1.59$ & $1.28\times 10^{-3}$  & $(1.24\pm0.29)10^{-3}$ & $-0.55$ \\
$\Lambda_c^+\to n\pi^+$ & $1.64$ & $-1.15$ & $0.50$ & $-3.45$ & $2.93$ & $-0.52$ & $$ & -- & $$ \\
$\Lambda_c^+\to \Lambda K^+$ & $1.66$ & $-0.08$ & $1.58$ & $-4.43$ & $0.55$ & $-3.70$ & $1.07\times 10^{-3}$ & $(6.1\pm1.2)10^{-4}$ & $-0.96$ \\
$\Lambda_c^+\to \Sigma^0 K^+$ & 0 & $1.49$ & $1.49$ &0 & $-2.29$  & $-2.29$ & $7.23\times 10^{-4}$ & $(5.2\pm0.8)10^{-4}$ & $-0.73$\\
$\Lambda_c^+\to \Sigma^+ K^0$ & 0 & $2.10$ & $2.10$ &0 & $-3.24$  & $-3.24$ & $1.44\times 10^{-3}$ & -- & $-0.73$\\
\end{tabular}
\end{ruledtabular}
\end{center}
}
\end{table}

We next turn to the mode $\Lambda_c^+\to\Xi^0 K^+$ which deserves a special attention. It has been shown that its $S$- and $P$-wave amplitudes are very small due to strong cancellation between various contributions. More specifically (see e.g. \cite{Cheng:1993gf}),
\be \label{eq:LambdacXiK}
A^{\rm{com}}(\Lambda_c^+\to \Xi^0 K^+) &=& \frac{1}{f_{K}}(a_{\Sigma^{+} \Lambda_{c}^{+}}-a_{\Xi^{0}\Xi_{c}^{0}}), \nonumber \\
B^{\rm{ca}}(\Lambda_c^+\to \Xi^0 K^+) &=&\frac{1}{f_{K}}\Bigg(g^{A({K}^{+})}_{\Xi^{0}\Sigma^{+}}\frac{m_{\Xi^{0}}+m_{\Sigma^{+}}}
{m_{\Lambda^{+}_{c}}-m_{\Sigma^{+}}}a_{\Sigma^{+}\Lambda^{+}_{c}} + a_{\Xi^{0}\Xi^{0}_{c}}\frac{m_{\Xi_c^{0}}+m_{\Lambda_c^{+}}}
{m_{\Xi^{0}}-m_{\Xi_c^{0}}}g^{A({K}^{+})}_{\Xi_c^{0}\Lambda_c^{+}}  \nonumber \\
&+& a_{\Xi^{0}\Xi^{'0}_{c}}\frac{m_{\Xi_c^{'0}}+m_{\Lambda_c^{+}}}
{m_{\Xi^{0}}-m_{\Xi_c^{'0}}}g^{A({K}^{+})}_{\Xi_c^{'0}\Lambda_c^{+}}
\Bigg).
\en
Since the matrix elements $a_{\Sigma^{+} \Lambda_{c}^{+}}$ and $a_{\Xi^{0}\Xi_{c}^{0}}$ are identical in the SU(3) limit and since there is a large cancellation between the first and third terms in $B^{\rm ca}$ (no contribution from the second term due to the vanishing  $g^{A({K}^{+})}_{\Xi_c^{0}\Lambda_c^{+}}$; for details see \cite{Cheng:1993gf}), the calculated branching fraction turns out to be too small compared to experiment and
the decay asymmetry is predicted to be zero owing to the vanishing $S$-wave amplitude \cite{Korner:1992wi,Xu:1992vc,Ivanov:1997ra,Zenczykowski:1993jm,Sharma:1998rd}.
This is a long-standing puzzle.

\begin{figure}[t]
\begin{center}
\includegraphics[width=0.90\textwidth]{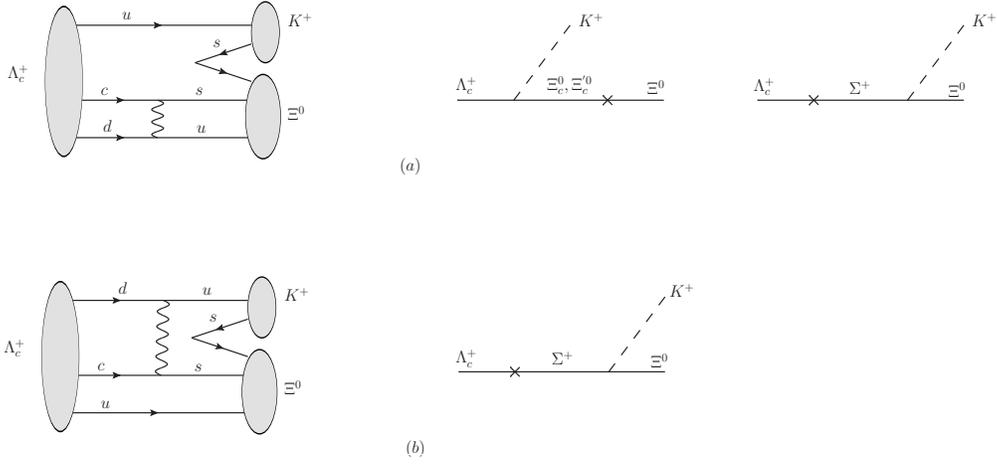}
\vspace{0.5cm}
\caption{$W$-exchange diagrams contributing to $\Lambda_c^+\to \Xi^0K^+$. The corresponding pole diagrams are also shown.} \label{fig:Lambdac_XiK} \end{center}
\end{figure}

To solve the above-mentioned puzzle, we notice that
one of the $W$-exchange diagrams depicted in the left panel of Fig. \ref{fig:Lambdac_XiK}(a) can be described by two distinct pole diagrams at the hadron level shown in the right panel of the diagram \ref{fig:Lambdac_XiK}(a).
These two pole diagrams are called type-III diagrams in \cite{Korner:1992wi} and
(d1) and (d2) in \cite{Zenczykowski:1993jm}. As first pointed out by K\"orner and Kr\"amer \cite{Korner:1992wi}, type-III diagram contributes only to the $P$-wave amplitude. Moreover, they pointed out that this diagram is empirically observed to be strongly suppressed.
It was argued by \.{Z}enczykowski
\cite{Zenczykowski:1993jm} that contributions from diagrams (d1) and (d2) cancel each other due to the  spin-flavor structure. Hence, its $S$- and $P$-wave amplitudes vanish.
The smallness of type-III $W$-exchange diagram also can be numerically checked through
Eq. (\ref{eq:LambdacXiK}).
In other words, the conventional expression of parity-violating and -conserving amplitudes given in Eq. (\ref{eq:LambdacXiK}) is actually for the type-III $W$-exchange diagram in Fig. \ref{fig:Lambdac_XiK}(a).
As a result, non-vanishing nonfactorizable $S$- and $P$-wave amplitudes arise solely from the $W$-exchange diagram depicted in Fig. \ref{fig:Lambdac_XiK}(b) (called type-II $W$-exchange diagram in \cite{Korner:1992wi} and (b)-type diagram in \cite{Zenczykowski:1993jm}).
The nonfactorizable amplitudes induced from type-II $W$-exchange now read
\be \label{eq:LambdacXiK_mod}
A^{\rm{com}}(\Lambda_c^+\to \Xi^0 K^+) &=& \frac{1}{f_{K}}a_{\Sigma^{+} \Lambda_{c}^{+}}, \nonumber \\
B^{\rm{ca}}(\Lambda_c^+\to \Xi^0 K^+) &=&\frac{1}{f_{K}}\left(g^{A({K}^{+})}_{\Xi^{0}\Sigma^{+}}\frac{m_{\Xi^{0}}+m_{\Sigma^{+}}}
{m_{\Lambda^{+}_{c}}-m_{\Sigma^{+}}}a_{\Sigma^{+}\Lambda^{+}_{c}}\right).
\en
Consequently, both partial wave amplitudes are not subject to large cancellations.

Note that the pole diagram induced by type-II $W$-exchange is the same as the second pole diagram (i.e. a weak transition of $\Lambda_c^+$--$\Sigma^+$  followed by a strong emission of $K^+$) in Fig. \ref{fig:Lambdac_XiK}(a), but it is no longer canceled by the first pole diagram. A vanishing $S$-wave amplitude was often claimed in the literature. We wish to stress again that the parity-violating amplitude can be induced from                                       type-II $W$-exchange through current algebra. \footnote{It had been argued that the contribution from type-II diagrams to the $S$-wave amplitude of $\Lambda_c^+\to\Xi^0K^+$ vanishes based on SU(4) symmetry \cite{Korner:1992wi,Zenczykowski:1993jm}. This is no longer true in the presence of SU(4)-symmetry breaking.}
Eq. (\ref{eq:LambdacXiK_mod}) leads to
${\cal B}(\Lambda_c^+\to \Xi^0 K^+)=0.71\%$, which is consistent with the data of $(0.55\pm0.07)\%$ \cite{Tanabashi:2018oca}.  Moreover, the predicted positive decay asymmetry of order $0.90$ is consistent with the value of $\alpha(\Lambda_c^+\to \Xi^0 K^+)=0.94^{+0.06}_{-0.11}$ obtained in the SU(3)-flavor approach \cite{Geng:2019xbo}.
\footnote{By measuring the angular dependence $1+\alpha_{\Xi K}\cos^2\theta_K$ in the process $\Lambda_c^+\to\Xi^0 K^+$, BESIII obtained $\alpha_{\Xi K}=0.77\pm0.78$ \cite{Ablikim:2018bir}. However, this quantity should not be confused with the decay asymmetry $\alpha(\Lambda_c^+\to \Xi^0 K^+)$ which is yet to be measured.
}
Therefore, the long-standing puzzle with the branching fraction and the decay asymmetry of $\Lambda_c^+\to\Xi^0 K^+$ is resolved.

In the $\Xi_c$ sector, vanishing type-III $W$-exchange contributions also occur in the CF decay $\Xi_c^0\to \Sigma^+ K^-$ and the SCS modes $\Xi_c^0\to pK^-,\Sigma^+\pi^-$. We will come to this point later.

Comparing Table \ref{tab:LambdacCF} with Table II of \cite{Cheng:2018hwl} for SCS $\Lambda_c^+$ decays, we see some changes in the $P$-wave amplitudes of $\Lambda_c^+\to p\pi^0,p\eta,n\pi^+$. This is because the first equation in (C2) of \cite{Cheng:2018hwl} should read
\be
g_{np}^{A(\pi^+)}=2g_{pp}^{A(\pi^0)}={10\over\sqrt{3}}g_{pp}^{A(\eta_8)}={5\over 3}(4\pi Z_1).
\en
Consequently, we find ${\cal B}(\Lambda_c^+\to p\pi^0)$ is modified from $0.75\times 10^{-4}$ \cite{Cheng:2018hwl} to the current value of $1.26\times 10^{-4}$. As for $\Lambda_c^+\to n\pi^+$, after correcting the error with the axial-vector form factor $g^{A(\pi^+)}_{\Sigma_c^0\Lambda_c}$ we find large cancellation in both $S$- and $P$-wave amplitudes, resulting a very small branching fraction of order $0.9\times 10^{-4}$. Since the large cancellation renders the present theoretical predictions of $\Lambda_c^+\to n\pi^+$ unreliable, we will not show its branching fraction and decay asymmetry in Tables \ref{tab:LambdacCF} and \ref{tab:compLambdac}.

\subsection{$\Xi_c$ decays}
Branching fractions and up-down decay asymmetries of CF and SCS $\Xi_c^{+,0}$ weak decays
are calculated according to Eqs. (\ref{eq:Gamma}),
(\ref{eq:alpha}) and  (\ref{eq:amplitude}), yielding the numerical results shown
in Tables \ref{tab:XiCF} and \ref{tab:SCS}, respectively.
One interesting point is that there does not exist any decay mode which proceeds only through the factorizable diagram.
Among all the processes, the three modes $\Xi_c^0\to\Sigma^+ K^-, \Xi^0\pi^0, \Xi^0\eta_8$ in CF processes
and the five SCS modes $\Xi_c^+\to p\overline{K}^0$, $\Xi_c^0\to\Xi^0 K^0, p K^-, n\overline{K}^0, \Sigma^+ \pi^-$
proceed only through the nonfactorizable diagrams,
while all the other channels receive contributions from both factorizable and nonfactorizable terms.
The relative sign between factorizable and nonfactorizable contributions determines whether the interference term is destructive or constructive.
For example, factorizable and nonfactorizable terms in both the $S$- and $P$-wave amplitudes of  the decays $\Xi_c^+\to \Sigma^+ \overline{K}^0, \Xi^0\pi^+$ and $\Xi_c^0\to \Sigma^0\overline{K}^0$
interfere destructively, leading to small branching fractions, especially for the last mode.
On the contrary, interference in the channels $\Xi_c^0\to \Lambda\overline{K}^0, \Xi^- \pi^+$ is found to be constructive.

The CF decay $\Xi_c^0\to \Sigma^+ K^-$ and the SCS modes $\Xi_c^0\to pK^-,\Sigma^+\pi^-$ are of special interest among all the $\Xi_c$ weak decays. Their naive $S$-wave amplitudes are given by
\be \label{eq:AcomXic}
A^{\rm{com}}(\Xi_c^0\to \Sigma^+ K^-) &=& \frac{1}{f_K} \left(a_{\Xi^0 \Xi_c^0}-a_{\Sigma^+\Lambda_c^+}\right),     \nonumber\\
A^{\rm{com}}(\Xi_c^0\to p K^-) &=& -\frac{1}{f_K}
\left(\frac{\sqrt{2}}{2} a_{\Sigma^0 \Xi_c^0} +\frac{\sqrt{6}}{2} a_{\Lambda \Xi_c^0}+a_{p\Lambda_c^+}\right),  \\
A^{\rm{com}}(\Xi_c^0\to \Sigma^+ \pi^-) &=& - \frac{1}{f_\pi} \left(\sqrt{2} a_{\Sigma^0 \Xi_c^0}+a_{\Sigma^+ \Xi_c^+}\right). \nonumber
\en
From Eqs. (\ref{eq:aCF}) and (\ref{eq:aSCS}) for baryon matrix elements, it is easily seen that they all vanish in the SU(3) limit. Likewise, their $P$-wave amplitudes are also subject to large cancellations.
Just as the decay $\Lambda_c^+\to\Xi^0K^+$ discussed in Sec. III.A, we should neglect the contributions from  type-III $W$-exchange diagrams and focus on type-II $W$-exchange ones.
The resulting amplitudes for these three modes now read
\be
A^{\rm{com}}(\Xi_c^0\to \Sigma^+ K^-) &=& \frac{1}{f_K}a_{\Xi^0 \Xi_c^0},     \nonumber\\
A^{\rm{com}}(\Xi_c^0\to p K^-) &=& -\frac{1}{f_K}
\left(\frac{\sqrt{2}}{2} a_{\Sigma^0 \Xi_c^0} +\frac{\sqrt{6}}{2} a_{\Lambda \Xi_c^0}\right),  \nonumber \\
A^{\rm{com}}(\Xi_c^0\to \Sigma^+ \pi^-) &=& - \frac{\sqrt{2}}{f_\pi} a_{\Sigma^0 \Xi_c^0}, \nonumber
\en
for $S$-wave (see Eqs. (\ref{eq:CFAcom}) and (\ref{eq:SCS0Acom})) and Eqs. (\ref{eq:CFBca}), (\ref{eq:SCS0Bca}) for $P$-wave. From Tables \ref{tab:XiCF} and \ref{tab:SCS} we see that
\be
\alpha(\Xi^0_c\to \Sigma^+K^-)\approx 0.98, \quad \alpha(\Xi^0_c\to pK^-)\approx0.99, \quad \alpha(\Xi^0_c\to \Sigma^+\pi^-)\approx 0.89\,.
\en
Hence, their decay asymmetries are all positive and close to unity. It is interesting to notice that the decay asymmetries of these three modes are also predicted to be positive and large in the SU(3) approach of \cite{Geng:2019xbo}.

\begin{table}[t]
\footnotesize{
\caption{ The Cabibbo-favored decays $\Xi_{c}\to \mathcal{B}_f P$ in units of  $10^{-2}G_F  {\rm{GeV}}^2$.
Branching fractions (in percent) and the up-down decay asymmetry $\alpha$ in theory and experiment are shown in the last four columns. Experimental results are taken from \cite{Li:2018qak,Li:2019atu,Edwards:1995xw} for branching fractions and \cite{Chan:2000kg} for decay asymmetry.
} \label{tab:XiCF}
\vspace{0.3cm}
\begin{center}
\begin{ruledtabular}
\begin{tabular}
{ l cccccc|cc cc}
 Channel & $A^{\rm{fac}}$ &  $A^{\rm{com}}$ & $A^{\rm{tot}}$ & $B^{\rm{fac}}$ &  $B^{\rm{ca}}$ & $B^{\rm{tot}}$ & $\mathcal{B}_{\rm{theo}}$
 & $\mathcal{B}_{\rm{exp}}$ &  $\alpha_{\rm{theo}}$&  $\alpha_{\rm{exp}}$\\
 \hline
 \colrule
$\Xi_{c}^{+}\to\Sigma^{+} \overline{K}^0$ & $2.98$  & $-4.48$  &$-1.50$    & $-9.95$  &   $12.28$   &
$2.32$  & $0.20$ & $-$  & $-0.80$   & $-$  \\
$\Xi_c^+\to\Xi^0\pi^+$  & $-7.41$  & $5.36$  &$-2.05$    & $28.07$  &   $-14.03$   &
$14.04$  & $1.72$ & $1.57\pm0.83$  & $-0.78$    & $-$ \\
\hline
$\Xi_c^0\to\Lambda\overline{K}^0$ &  $-1.11$  & $-5.41$  &$-6.52$    & $3.66$  &   $6.87$   &
$10.52$  & $1.33$ & $-$  & $-0.86$   & $-$  \\
$\Xi_c^0\to\Sigma^0\overline{K}^0$ &  $-2.11$  & $3.12$  &$1.02$    & $7.05$  &   $-9.39$   &
$-2.33$  & $0.04$ & $-$  & $-0.96$ & $-$    \\
$\Xi_c^0\to\Sigma^+ K^-$ & $0$  & $-4.42$  &$-4.42$    & $0$  &   $-12.09$~~   &
$-12.09$~~  & $0.46$ & $-$  & ~~$0.18$   & $-$\\
$\Xi_c^0\to\Xi^0\pi^0$ & $0$  & $-7.58$  &$-7.58$    & $0$  &   $11.79$   &
$11.79$  & $1.82$ & $-$  & $-0.77$   & $-$\\
$\Xi_c^0\to\Xi^0\eta$& $0$  & $-10.80$  &$-10.80$    & $0$  &   $-6.17$   &
$-6.17$  & $2.67$ & $-$  & ~~$0.30$  & $-$ \\
$\Xi_c^0\to\Xi^-\pi^+$  &  $-7.42$  & $-5.36$  &$-12.78$    & $28.24$  &   $2.65$   &
$30.89$  & $6.47$ & $ 1.80\pm 0.52$  & $-0.95$  & $-0.6\pm 0.4$   \\
\end{tabular}
\end{ruledtabular}
\end{center}
}
\end{table}

Besides the above-mentioned three modes of $\Xi_c^0$, type-III $W$-exchange diagram also exists in the following channels: $\Xi_c^+\to \Xi^0K^+,\Sigma^+(\pi^0,\eta)$ and
$\Xi_c^0\to (\Lambda,\Sigma^0)\pi^0, (\Lambda,\Sigma^0)\eta$. However, the effects of vanishing
type-III $W$-exchange can be seen only in the $P$-wave amplitudes of $\Xi_c^+\to \Sigma^+\pi^0$ and $\Xi_c^0\to (\Lambda,\Sigma^0)\pi^0$. In Eqs. (\ref{eq:SCSBca}) and (\ref{eq:SCS0Bca}) for these three modes we have explicitly dropped the pole contributions with the strong $\pi^0$ emission from $\Xi_c$ followed by a weak transition.

As for the two modes $\Xi_c^+\to\Xi^0\pi^+$ and $\Xi_c^0\to\Xi^-\pi^+$, we see from Table \ref{tab:XiCF} that our prediction is in good agreement with experiment for the former, but it is too large compared to the experimental measurement for the latter.
This is mainly due to the relative sign between factorizable
and nonfactorizable terms. In the absence of nonfactorizable contributions, we find ${\cal B}(
\Xi_c^+\to\Xi^0\pi^+)\approx 9.9\%$ and ${\cal B}(\Xi_c^0\to\Xi^-\pi^+)\approx 3.3\%$. Since the measured branching fractions are $(1.57\pm0.83)\%$ and $(1.80\pm0.52)\%$, respectively, this implies that there should be a large destructive interference between factorizable and nonfactorizable terms in the former and a smaller  destructive interference in the latter.
We notice that the factorizable amplitudes of these two modes are very similar.
\footnote{We have confirmed that the sign of the factorizable contribution in the earlier work of \cite{Cheng:1993gf} has to be flipped due to the sign convention with the form factors $f_1$ and $g_1$.}
From Eq. (\ref{eq:CFAcom}), it is clear that the commutator terms of both modes denoted by $A^{\rm com}$ are the same in magnitude but opposite in sign. Consequently, the interference between $A^{\rm fact}$ and $A^{\rm com}$ is destructive in $\Xi_c^+\to\Xi^0\pi^+$ but constructive in  $\Xi_c^0\to\Xi^-\pi^+$ (see also \cite{Xu:1992vc}). As a result, the predicted branching fraction of order $6.5\%$ for the latter is too large.
If we use the form factors $f_1^{\Xi_c\Xi}(0)=-0.590$ and $g_1^{\Xi_c\Xi}(0)=-0.582$ \cite{Faustov:2019ddj} in conjunction with the $q^2$ dependence given by Eq. (\ref{eq:FF1}), the branching fraction will be reduced only slightly from $6.5\%$ to $6.2\%$. Hence, we conclude that these two modes cannot be simultaneously explained within the framework of current algebra for $S$-wave amplitudes.

\begin{table}[t]
\linespread{1.5}
\caption{ The singly Cabibbo-suppressed decays $\Xi_{c}\to \mathcal{B}_f P$ in units of  {$10^{-2}G_F  {\rm{GeV}}^2$}.
 Branching fractions (in unit of $10^{-3}$) and the asymmetry parameter $\alpha$ are shown in the last two columns.
} \label{tab:SCS}
\begin{center}
\begin{ruledtabular}
\begin{tabular}
{ l rrrrrr|r r}
 Channel & $A^{\rm{fac}}$ &  $A^{\rm{com}}$ & $A^{\rm{tot}}$ & $B^{\rm{fac}}$ &  $B^{\rm{ca}}$ & $B^{\rm{tot}}$~~ & $\mathcal{B}_{\rm{theo}}$
 &  $\alpha_{\rm{theo}}$\\
 \hline
  \colrule
  $\Xi_c^+\to\Lambda\pi^+$ &  $0.46$  & $-1.50$  &$-1.04$    & $-1.69$  &   $2.16$   &
$0.47$~~  & $0.85$  & $-0.33$     \\
$\Xi_c^+\to\Sigma^0\pi^+$ &  $-0.90$  & $-1.00$  &$-1.90$    & $3.29$  &   $0.74$   &
$4.03$~~ & $4.30$  & $-0.95$     \\
$\Xi_c^+\to\Sigma^+\pi^0$ &  $0.32$  & $1.00$  &$1.32$    & $-1.16$  &   $1.61$   &
$0.44$~~  & $1.36$   & $0.23$     \\
$\Xi_c^+\to\Sigma^+\eta$ &  $-0.74$  & $1.42$  &$0.68$    & $2.58$  &   $-2.19$   &
$0.39$~~  & $0.32$ & $0.36$     \\
$\Xi_c^+\to p\overline{K}^0$  &  $0$  & $-2.10$  &$-2.10$    & $0$  &   $2.64$   &
$2.64$~~  & $3.96$  & $-0.83$ \\
$\Xi_c^+\to\Xi^0 K^+$ &  $-2.30$  & $1.16$  &$-1.14$    & $8.43$  &   $-3.46$   &
$4.97$~~  & $2.20$  & $-0.98$     \\
\hline
$\Xi_c^0\to\Lambda\pi^0$ &  $-0.12$  & $1.06$  &$0.95$    & $0.42$  &   $-0.96$   &
$-0.53$~~  & $0.24$   & $-0.41$     \\
$\Xi_c^0\to\Lambda\eta$ & $0.27$  & $1.51$  &$1.78$    & $-0.94$  &   $-0.71$   &
$-1.65$~~  & $0.81$  & $-0.59$     \\
$\Xi_c^0\to\Sigma^0\pi^0$&   $-0.23$  & $-0.70$  &$-0.93$    & $0.82$  &   $1.36$   &
$2.18$~~  & $0.38$  & $-0.98$     \\
$\Xi_c^0\to\Sigma^0\eta$& $0.53$  & $-1.01$  &$-0.48$    & $-1.83$  &   $1.55$   &
$-0.28$~~  & $0.05$  & $0.36$     \\
$\Xi_c^0\to\Sigma^-\pi^+$ &   $-1.28$  & $-1.41$  &$-2.69$    & $4.67$  &   $0.22$   &
$4.89$~~  & $2.62$  & $-0.90$     \\
$\Xi_c^0\to\Sigma^+ \pi^-$ & $0$  & $1.41$  &$1.41$    & $0$  &   $2.49$   &
$2.49$~~  & $0.71$   & $0.89$ \\
$\Xi_c^0\to p K^-$ & $0$  & $-0.94$  &$-0.94$    & $0$  &   $-1.86$   &
$-1.86$~~  & $0.35$   & $0.99$   \\
$\Xi_c^0\to n \overline{K}^0$ & $0$  & $-2.10$  &$-2.10$    & $0$  &   $2.96$   &
$2.96$~~  & $1.40$   & $-0.89$   \\
$\Xi_c^0\to\Xi^0 K^0$ & $0$  & $2.10$  &$2.10$    & $0$  &   $-4.17$   &
$-4.17$~~  & $1.32$  & $-0.85$   \\
$\Xi_c^0\to\Xi^- K^+$ &  $-2.31$  & $-0.94$  &$-3.24$    & $8.49$  &   $0.71$   &
$9.20$~~  & $3.90$   & $-0.97$     \\
\end{tabular}
\end{ruledtabular}
\end{center}
\end{table}

To circumvent the difficulty with $\Xi_c^0\to\Xi^-\pi^+$, one possibility is to consider the correction to the current-algebra calculation of the parity-violating amplitude by writing
\begin{equation}
A=A^{\rm{CA}}+(A-A^{\rm{CA}}),
\end{equation}
where the term $(A-A^{\rm{CA}})$ can be regarded as an on-shell correction to the current-algebra ressult. It turns out that in the existing pole model calculations
\cite{Cheng:1991sn, Cheng:1993gf,Xu:1992vc}, the on-shell correction  $(A-A^{\rm CA})$ always has a sign opposite to that of $A^{\rm CA}$. Moreover, the on-shell correction is sometimes large enough to flip the sign of the parity-violating amplitudes. It is conceivable that on-shell corrections could be large for $\Xi^-\pi^+$ but small for $\Xi^0\pi^+$. This issue needs to be clarified in the future. Nevertheless, we have learned from Table \ref{tab:LambdacCF} that current algebra generally works well in $\Lambda_c^+\to {\cal B}+P$ decays,


For the up-down decay asymmetry, there is only one measurement thus far.
In 2001, CLEO collaboration measured $\Xi_c^0\to\Xi^-\pi^+$ and found
$\alpha({\Xi_c^0\to\Xi^-\pi^+})=-0.6\pm 0.4$ \cite{Chan:2000kg}.
Our prediction 
is consistent with the CLEO's value.
Decay asymmetries are usually negative in most of the channels. However,
besides the three modes $\Xi_c^0\to \Sigma^+K^-,pK^-,\Sigma^+\pi^-$ as discussed before, the following channels
$\Xi_c^0\to \Xi^0\eta, \Sigma^0\eta$ and $\Xi_c^+\to \Sigma^+\pi^0, \Sigma^+\eta$ in the $\Xi_c$ sector
are also predicted to have positive decay asymmetries (see Tables \ref{tab:XiCF} and \ref{tab:SCS}).
We hope that these predictions could be tested in the near future by Belle/Belle II.

\subsection{Comparison with the SU(3) approach}

Besides dynamical model calculations, two-body nonleptonic decays of charmed baryons have been analyzed in terms of SU(3)-irreducible-representation
amplitudes \cite{Savage,Verma}. There are two distinct  approaches to implement this idea. One is to write down the SU(3)-irreducible-representation
amplitudes by decomposing the effective Hamiltonian through the Wigner-Eckart theorem. The other is to use the topological quark diagrams which are related in different decay channels via SU(3) flavor symmetry. Each approach has its own advantage.   A general formulation of the quark-diagram scheme for charmed baryons is given in \cite{CCT} (see also \cite{Kohara}).  Analysis of  Cabibbo-suppressed decays using SU(3) flavor symmetry was first carried out in \cite{Sharma}. This approach became very popular recently \cite{Lu,Geng:2018plk,Geng:2018bow,Geng:2019xbo}.  Although SU(3) flavor symmetry is approximate, it does provide very useful information. In Tables \ref{tab:compLambdac}
and \ref{tab:compXic} we
compare our results for $\Lambda_c^+$ and $\Xi_c^{+,0}$ decays, respectively, with  the $SU(3)_F$ approach in \cite{Geng:privite,Geng:2019xbo} in which the parameters for both $S$- and $P$-wave amplitudes are obtained by fitting to the data.
\footnote{Many early studies in the $SU(3)_F$ approach have overlooked the fact that charmed baryon decays are governed by several different partial-wave amplitudes which have distinct kinematic and dynamic effects.}

\begin{table}[tp!]
 \linespread{1.3}
\begin{ruledtabular}
\linespread{1.3}
 \caption{ Comparison of this work with \cite{Geng:privite,Geng:2019xbo} for
 the branching fractions in units of $10^{-2}$ for Cabibbo-favored $\Lambda_c^+$ decays (upper entry) and $10^{-3}$ for singly Cabibbo-suppressed ones (lower entry). Decay asymmetries are shown in parentheses.
 } \label{tab:compLambdac}
\begin{center}
\begin{tabular}
{ l *4{c}}
 Modes  & This work & Geng {\it et al.} \cite{Geng:privite,Geng:2019xbo}
 & Expt.\\
 \hline
$\Lambda_c^+\to\Lambda\pi^+$  & $1.30$ ($-0.93$)   & $1.27\pm 0.07$ ($-0.77\pm0.07$)    & $1.30\pm0.07$ ($-0.84\pm0.09$) \\
$\Lambda_c^+\to\Sigma^0\pi^+$  & $2.24$ ($-0.76$)   & $1.26\pm 0.06$ ($-0.58\pm0.10$)    & $1.29\pm0.07$ ($-0.73\pm0.18$) \\
$\Lambda_c^+\to\Sigma^+\pi^0$  & $2.24$ ($-0.76$)   & $1.26\pm 0.06$ ($-0.58\pm0.10$)    & $1.25\pm0.10$ ($-0.55\pm0.11$) \\
$\Lambda_c^+\to\Sigma^+\eta$  & $0.74$ ($-0.95$)   & $0.29\pm 0.12$~~ ($-0.70^{+0.59}_{-0.30}$)    & $0.53\pm0.15$ \\
$\Lambda_{c}^{+}\to p \overline{K}^0$ & $2.11$ ($-0.75$)   & $3.14\pm0.15$~~ ($-0.99^{+0.09}_{-0.01}$)  &  $3.18\pm0.16$~ ($~0.18\pm0.45$)   \\
$\Lambda_c^+\to\Xi^0 K^+$  & $0.73$~ ($~0.90$)   & $0.57\pm 0.09$~~~~ ($1.00^{+0.00}_{-0.02}$)    & $0.55\pm0.07$~ ($~0.77\pm0.78$) \\
\hline
$\Lambda_c^+\to p\pi^0$  & $0.13$ ($-0.97$)   & $0.11^{+0.13}_{-0.11}$~~~~ ($~0.24\pm0.68$)  & $<0.27$  \\
$\Lambda_c^+\to p\eta$  & $1.28$ ($-0.55$)   & $1.12\pm 0.28$~~ ($-1.00^{+0.06}_{-0.00}$)  & $1.24\pm0.29$  \\
$\Lambda_c^+\to n\pi^+$  &   & $0.76\pm 0.11$~ ($~0.27\pm0.11$)  &  \\
$\Lambda_c^+\to \Lambda K^+$  & $1.07$ ($-0.96$)   & $0.66\pm 0.09$~ ($~0.09\pm0.26$)  & $0.61\pm0.12$  \\
$\Lambda_c^+\to \Sigma^0 K^+$  & $0.72$ ($-0.73$)   & $0.52\pm 0.07$~~ ($-0.98^{+0.05}_{-0.02}$)  & $0.52\pm0.08$  \\
$\Lambda_c^+\to \Sigma^+ K^0$  & $1.44$ ($-0.73$)   & $1.05\pm 0.14$~~ ($-0.98^{+0.05}_{-0.02}$)  & $$  \\
\end{tabular}
\end{center}
\end{ruledtabular}
\end{table}

We see from Table \ref{tab:compLambdac} that it appears the SU(3) approach gives a better
description of the measured branching fractions because it fits to the data. However, it is worth of mentioning that in the beginning the SU(3) practitioners tended to make the assumption of the sextet {\bf 6} dominance over $\overline{\bf 15}$. Under this hypothesis,
one will lead to ${\cal B}(\Lambda_c^+\to p\pi^0)\sim 5\times 10^{-4}$ \cite{Lu,Geng:2018plk}, which exceeds the current experimental limit of $2.7\times 10^{-4}$ \cite{BES:peta}. Our dynamic calculation in \cite{Cheng:2018hwl} predicted ${\cal B}(\Lambda_c^+\to p\pi^0)\sim 1\times 10^{-4}$. As far as the branching fraction is concerned, it is important to measure the mode $\Lambda_c^+\to n\pi^+$ to distinguish our prediction from the SU(3) approach.
As for decay asymmetries, while we agree on the sign and magnitude of $\alpha(\Xi^0 K^+)$, we disagree on the sign of $\alpha$ in $\Lambda K^+$. Hopefully, these can be tested in the future.

It is clear from Table \ref{tab:compXic} that  except $\Xi_c^+\to\Sigma^+\overline{K}^0, \Xi^0\pi^+,\Xi^0K^+$ and $\Xi_c^0\to\Xi^-\pi^+,\Xi^-K^+$ all the branching fractions of $\Xi_c^{+,0}$ decays in this work and in the SU(3) approach are consistent with each other within a factor of 2. Furthermore, we agree on the signs of decay asymmetries except $\Xi_c^+\to\Sigma^+\overline{K}^0$ and $\Xi_c^+\to\Xi^0K^+$.
\footnote{Those predictions of $\alpha$ with the uncertainty greater than the central value are not taken into account for comparison.}
Notice that both approaches lead to ${\cal B}(\Xi_c^0\to\Xi^-\pi^+)\gg {\cal B}(\Xi_c^+\to\Xi^0\pi^+)$, contrary to the current data. Hence, it is of great importance to measure the branching fractions of them more accurately in order to test their underlying mechanism.

\begin{table}[tp!]
 \linespread{1.7}
\begin{ruledtabular}
\linespread{1.3}
 \caption{ Comparison of this work with \cite{Geng:privite,Geng:2019xbo} for
 the branching fractions in units of $10^{-2}$ for Cabibbo-favored $\Xi_c^{+,0}$ decays (upper entry) and $10^{-3}$ for singly Cabibbo-suppressed ones (lower entry). Decay asymmetries are shown in parentheses. Experimental results are taken from \cite{Li:2018qak,Li:2019atu,Edwards:1995xw} for branching fractions and \cite{Chan:2000kg} for decay asymmetry.
 } \label{tab:compXic}
\begin{center}
\begin{tabular}
{ l *3{c}}
 Modes  & This work & Geng \emph{et al.} \cite{Geng:privite,Geng:2019xbo}
 & Expt.\\
 \hline
$\Xi_{c}^{+}\to\Sigma^{+} \overline{K}^0$ & $0.20$ ($-0.80$)   & $0.78^{+1.02}_{-0.78}$~~~~~~~~ ($0.93^{+0.07}_{-0.14}$)   &    \\
$\Xi_c^+\to\Xi^0\pi^+$  & $1.72$ ($-0.78$)   & $0.42\pm 0.17$ ($-0.43\pm0.57$)    & $1.57\pm 0.83$ \\
$\Xi_c^0\to\Lambda\overline{K}^0$ &  $1.33$ ($-0.86$)  & $1.42\pm 0.09$~~~ ($-0.85^{+0.16}_{-0.15}$)  &    \\
$\Xi_c^0\to\Sigma^0\overline{K}^0$ &  $0.04$ ($-0.94$)  & $0.09^{+0.11}_{-0.09}$~~~~~ ~~~($0.30^{+0.70}_{-0.84}$)  &   \\
$\Xi_c^0\to\Sigma^+ K^-$ & $0.46$~ ($~0.18$)    &$0.76\pm0.14$~~~~ ($~0.93^{+0.07}_{-0.08}$)   &   \\
$\Xi_c^0\to\Xi^0\pi^0$ & $1.82$ ($-0.77$)   & $1.00\pm 0.14$~~~ ($-0.96^{+0.05}_{-0.04}$)   &   \\
$\Xi_c^0\to\Xi^0\eta$& $2.67$~ ($~0.30$)   &$1.30\pm 0.23$~ ($~0.80\pm0.16$)   &   \\
$\Xi_c^0\to\Xi^-\pi^+$  &  $6.47$ ($-0.95$)    & $2.95\pm 0.14$~ ~~($-1.00^{+0.01}_{-0.00}$) & $1.80\pm 0.52$ ($-0.6\pm0.4$)
 \\
 \hline
$\Xi_c^+\to\Lambda\pi^+$ &  $0.85$ ($-0.33$)    &$1.23\pm 0.42$~~ ($0.03\pm0.18$)   &  \\
$\Xi_c^+\to\Sigma^0\pi^+$ &  $4.30$ ($-0.95$)   & $2.65\pm 0.25$ ($-0.61\pm0.12$) &  \\
$\Xi_c^+\to\Sigma^+\pi^0$ &  $1.36$~ ($~0.23$)  & $2.61\pm 0.67$ ($-0.18\pm0.36$)    &   \\
$\Xi_c^+\to\Sigma^+\eta$ &  $0.32$~ ($~0.36$)  & $1.50\pm1.06$~~ ($~0.30\pm0.60$)     &    \\
$\Xi_c^+\to p\overline{K}^0$  &  $3.96$ ($-0.83$)  &  $4.64\pm 0.72$ ($-0.83\pm0.06$)  &   \\
$\Xi_c^+\to\Xi^0 K^+$ &  $2.20$ ($-0.98$)  & $0.76\pm 0.12$~ ($~0.39\pm0.16$)     &    \\
$\Xi_c^0\to\Lambda\pi^0$ &  $0.24$ ($-0.41$)  & $0.31\pm 0.11$~~ ($0.08\pm0.22$)     &      \\
$\Xi_c^0\to\Lambda\eta$ & $0.81$ ($-0.59$)  & $0.79\pm0.27$ ($-0.17\pm0.26$)     &     \\
$\Xi_c^0\to\Sigma^0\pi^0$&   $0.38$ ($-0.98$)  & $0.50\pm 0.09$ ($-0.74\pm0.25$)      &    \\
$\Xi_c^0\to\Sigma^0\eta$& $0.05$~ ($~0.36$)  &$0.18\pm0.11$ ($-0.20\pm0.76$)      &   \\
$\Xi_c^0\to\Sigma^-\pi^+$ &   $2.62$ ($-0.90$)  & $1.83\pm 0.09$ ($-0.99\pm0.01$)   &  \\
$\Xi_c^0\to\Sigma^+ \pi^-$ & $0.71$~ ($~0.89$)  & $0.49\pm 0.09$~ ($~0.91\pm0.09$)   &   \\
$\Xi_c^0\to p K^-$ & $0.35$~ ($~0.99$)  & $0.60\pm 0.13$~ ($~0.82\pm0.11$)    &     \\
$\Xi_c^0\to n \overline{K}^0$ & $1.40$ ($-0.89$) & $1.07\pm 0.06$ ($-0.74\pm0.12$)    &    \\
$\Xi_c^0\to\Xi^0 K^0$ & $1.32$ ($-0.85$)  & $0.96\pm 0.04$ ($-0.53\pm0.09$)       &    \\
$\Xi_c^0\to\Xi^- K^+$ &  $3.90$ ($-0.97$)  &$1.28\pm 0.06$~~~ ($-1.00^{+0.01}_{-0.00}$)   &     \\
\end{tabular}
\end{center}
 \end{ruledtabular}
 \end{table}

\subsection{Theoretical uncertainties}
In this subsection we discuss the major theoretical uncertainties one may encounter in this work. 

(i) Wave functions in the MIT bag model. In the bag model the quark spatial wave function in the ground $1S_{1/2}$ state has the expression
\be \label{eq:uv}
\psi_{1S_{1/2}}=
\left(
\begin{array}{c}
iu(r)\chi \\
v(r){\boldsymbol\sigma}\cdot{\bf \hat{r}}\chi \\
\end{array}
\right),
\en
where $u(r)$ and $v(r)$ are the large and small components of the quark wave function, respectively. In this work, we have employed the following bag parameters
\be
m_u=m_d=0, \quad m_s=0.279~{\rm GeV}, \quad m_c=1.551~{\rm GeV}, \quad R=5~{\rm GeV}^{-1},
\en
where $R$ is the radius of the bag.  The uncertainties in the bag parameters will affect the estimation of hadron matrix elements, form factors and the strong couplings.

(ii)  form factors and Wilson parameters in factorizable amplitudes.
The uncertainties in the factorizable amplitudes given in Eq. (\ref{eq:factSCS}) arise from the Wilson parameters $a_{1,2}$ and the form factors $f_1(m_P^2)$ and $g_1(m^2_P)$. The measurement of $\Lambda_c^+\to p\phi$ allows us to fix $a_2$ to be $-0.45\pm0.05$ \cite{Cheng:2018hwl} which in turn implies that $a_1=1.26\pm0.02$. Form factors are first evaluated at zero recoil using the bag model. Their $q^2$ dependence is then determined based on the assumption of nearest pole dominance.

(iii) Nonfactorizable $S$-wave amplitude in current algebra.  We have employed current algebra to evaluate $S$-wave amplitudes to circumvent the troublesome $1/2^-$ intermediate baryon resonances which are not well understood in the quark model. Since current algebra is valid in the soft meson limit, it is natural to expect an correction of order $q^2/\Lambda_\chi^2$ where $\Lambda_\chi\sim 1$ GeV is a chiral symmetry breaking scale and $q^2$ is the c.m. three-momentum squared of the pseudoscalar meson produced in charmed baryon decays.

Among the antitriplet charmed baryon decays, $\Lambda_c^+\to p\phi$ is the only purely factorizable process. Also it is very difficult to quantify the errors from part (iii). Therefore, we will focus on the uncertainties arising from the wave functions in the bag model. By varying the bag radius $R$ from 5.0 GeV$^{-1}$ (or 0.987 fm) to 4.8 and 5.2 GeV$^{-1}$, we obtain bag integrals slightly different from that given in Eqs. (\ref{eq:Xi}) and (\ref{eq:Zi}). This allows to estimate the uncertainties in baryon matrix elements and the axial-vector form factors. Take $\Lambda_c^+\to \Xi^0 K^+$ and $\Xi_c^0\to \Sigma^+K^-$ as examples for illustration as they proceed only through nonfactorizable diagrams.
We obtain ${\cal B}(\Lambda_c^+\to \Xi^0 K^+)=(0.73^{+0.20}_{-0.15})\%$ and ${\cal B}(\Xi_c^0\to \Sigma^+ K^-)=(0.46^{+0.13}_{-0.10})\%$. Hence a slight change of the bag radius by 4\% will result in $(20-30)\%$ uncertainties in branching fractions.

\section{Conclusion}
\label{sec:con}

In this work we have systematically studied the branching fractions and up-down decay asymmetries of CF and SCS  decays of antitriplet charmed baryons.
To estimate the nonfactorizable contributions, we work in the pole model for the $P$-wave amplitudes and current algebra for $S$-wave ones.
Throughout the whole calculations, all the non-perturbative parameters, including form factors, baryon matrix elements and
axial-vector form factors are evaluated using the MIT bag model.

We draw some conclusions from our analysis:
\begin{itemize}

\item The long-standing puzzle with the branching fraction and decay asymmetry of $\Lambda_c^+\to\Xi^0 K^+$ is resolved by realizing that only type-II $W$-exchange diagram will contribute to this mode. We find that not only the predicted rate agrees with experiment but also the  decay asymmetry is consistent in sign and magnitude with the SU(3) flavor approach. Hence, it is most likely that $\alpha(\Lambda_c^+\to\Xi^0 K^+)$ is large and positive.
\item
In analog to $\Lambda_c^+\to\Xi^0 K^+$, the CF mode $\Xi_c^0\to \Sigma^+K^-$ and the SCS decays $\Xi_c^0\to pK^-,\Sigma^+\pi^-$ proceed only through type-II $W$-exchange. They are predicted to have large and {\it positive} decay asymmetries. This can be tested in the near future.

\item
The predicted $\mathcal{B}(\Xi_c^+\to
\Xi^0\pi^+)$ agrees well with the measurement inferred from Belle and CLEO, while
the calculated ${\cal B}(\Xi_c^0\to \Xi^-\pi^+)$ is too large compared to the recent Belle
measurement. We find ${\cal B}(\Xi_c^0\to \Xi^-\pi^+)\gg \mathcal{B}(\Xi_c^+\to
\Xi^0\pi^+)$ and conclude that these two modes cannot be simultaneously explained within the current-algebra framework for $S$-wave amplitudes. On-shell corrections to the current-algebra results are probably needed to circumvent the difficulty with $\Xi_c^0\to \Xi^-\pi^+$. More accurate measurements of them are called for to set the issue.

\item Owing to large cancellation between factorizable and nonfactorizable contributions for both $S$- and $P$-wave amplitudes, we argue that the present theoretical predictions of $\Lambda_c^+\to n\pi^+$ are unreliable. It is important to measure this SCS mode to understand its underlying mechanism.

\item Although $\Xi_c^0\to \Sigma^0\overline{K}^0$ and $\Xi_c^+\to \Sigma^+\overline{K}^0$
are Cabibbo-favored decays, their branching fractions are small especially for the former due to large destructive interference between factorizable and nonfactorizable amplitudes.

\item We have compared our results with the approach of SU(3) flavor symmetry.  Excluding those predictions of $\alpha$ with the uncertainty greater than the central value, we find that both approaches agree on the signs of decay asymmetries except the three modes:  $\Lambda_c^+\to n\pi^+$, $\Xi_c^+\to\Sigma^+\overline{K}^0$ and $\Xi_c^+\to\Xi^0K^+$. We also agree on the hierarchy ${\cal B}(\Xi_c^0\to \Xi^-\pi^+)\gg \mathcal{B}(\Xi_c^+\to \Xi^0\pi^+)$.

\item We have identified several major sources of theoretical uncertainties and gave some crude estimation of errors on branching fractions provided that uncertainty arise from the MIT bag-model wave functions. 

\end{itemize}

\begin{acknowledgments}
We would like to thank C. Q. Geng for discussion and for providing us the updated results in the flavor-SU(3) approach.
This research was supported in part by the Ministry of Science and Technology of R.O.C. under Grant No. 107-2119-M-001-034. F. Xu is supported by NSFC under Grant Nos. 11605076 and U1932104.
\end{acknowledgments}

\appendix

\section{Baryon wave functions}
\label{app:wf}
Throughout this paper, we follow the convention in \cite{Cheng:2018hwl} for the wave functions of baryons with $S_z=1/2$:
\be
p &=& {1\over\sqrt{3}}\left[ uud\chi_{_S}+(13)+(23)\right], \qquad \qquad \quad ~~n = -{1\over\sqrt{3}}\left[ ddu\chi_{_S}+(13)+(23)\right], \non \\
\Sigma^+ &=& -{1\over\sqrt{3}}\left[ uus\chi_{_S}+(13)+(23)\right], \qquad\qquad \quad
\Sigma^0 = {1\over\sqrt{6}}\left[ (uds+dus)\chi_{_S}+(13)+(23)\right], \non \\
 \Xi^0&=& {1\over\sqrt{3}}\left[ ssu\chi_{_S}+(13)+(23)\right], \qquad \qquad\quad ~~ \Xi^- = {1\over\sqrt{3}}\left[ ssd\chi_{_S}+(13)+(23)\right],  \non \\
\Lambda &=& -{1\over\sqrt{6}}\left[ (uds-dus)\chi_{_A}+(13)+(23)\right], \quad
\Lambda_c^+ = -{1\over\sqrt{6}}\left[ (udc-duc)\chi_{_A}+(13)+(23)\right],  \non \\
\Sigma_c^+ &=& {1\over\sqrt{6}}\left[ (udc+duc)\chi_{_S}+(13)+(23)\right], \qquad
\Sigma_c^0 = {1\over\sqrt{3}}\left[ ddc\chi_{_S}+(13)+(23)\right], \label{eq:baryonwf} \\
\Xi_c^+ &=& {1\over\sqrt{6}}\left[ (usc-suc)\chi_{_A}+(13)+(23)\right], \qquad
\Xi_c^0 = {1\over\sqrt{6}}\left[ (dsc-sdc)\chi_{_A}+(13)+(23)\right], \non \\
\Xi_c^{'+} &=& {1\over\sqrt{6}}\left[ (usc+suc)\chi_{_S}+(13)+(23)\right], \qquad
\Xi_c^{'0} = {1\over\sqrt{6}}\left[ (dsc+sdc)\chi_{_S}+(13)+(23)\right], \non\\
\Omega_c^0 &=& \frac{1}{\sqrt{3}}\left[ssc\chi_s +(13)+(23)\right],\qquad\qquad \quad ~~
\Sigma^-=\frac{1}{\sqrt{3}}[dds \chi_s +(13)+(23)],   \non
\en
where $abc\chi_{_S}=(2a^\up b^\up c^\dw-a^\up b^\dw c^\up-a^\dw b^\up c^\up)/\sqrt{6}$ and $abc\chi_{_A}=(a^\up b^\dw c^\up-a^\dw b^\up c^\up)/\sqrt{2}$.
There are  two  useful relations under the $U$-,$V$-, and $I$-spin:
\begin{align}
&abc\chi_s +\sqrt{3} abc\chi_A = -\sqrt{\frac23} (
2a^{\downarrow} b^{\uparrow} c^{\uparrow}
-a^{\uparrow}b^{\uparrow}c^{\downarrow}-a^{\uparrow} b^{\downarrow} c^{\uparrow}), \nonumber\\
&abc\chi_s -\sqrt{3} abc\chi_A = -\sqrt{\frac23} (
2a^{\uparrow} b^{\downarrow} c^{\uparrow}
-a^{\uparrow}b^{\uparrow}c^{\downarrow}-a^{\downarrow} b^{\uparrow} c^{\uparrow}).
\end{align}

\section{Baryons under $U$-,$V$- and $I$-spin}
\label{app:UVI}

In practical calculations, we need to specify the behaviors of baryon wave functions under the isospin, $U$-spin and $V$-spin ladder operators. Based on  the wave functions given by Eq. (\ref{eq:baryonwf}), we have the following relations
\begin{eqnarray}
U_+|\Sigma^+\ra &= &-|p\ra ,\qquad\qquad\qquad\qquad\quad  U_+|\Xi^-\ra = -|\Sigma^-\ra,         \nonumber\\
U_+|\Sigma^0\ra &=& \frac{\sqrt{2}}{2} |n\ra,\qquad\qquad\qquad\qquad U_+ |\Lambda\ra=-\frac{\sqrt{6}}{2} |n\ra, \nonumber \\
U_+ |\Xi^0\ra &=& -\frac{\sqrt{2}}{2} |\Sigma^0\ra +\frac{\sqrt{6}}{2}|\Lambda\ra,\qquad~~
U_+ |\Xi_c^+\ra =-|\Lambda_c^+\ra,   \\
U_-|\Sigma^0\ra &=& -\frac{\sqrt{2}}{2} |\Xi^0\ra,\qquad\qquad\qquad\quad U_- |\Lambda\ra=\frac{\sqrt{6}}{2} |\Xi^0\ra,  \nonumber \\
U_- |n\ra &=& \frac{\sqrt{2}}{2} |\Sigma^0\ra -\frac{\sqrt{6}}{2}|\Lambda\ra,
\qquad\quad~~ {{ U_-|p\ra =-|\Sigma^+\ra, }}
\nonumber
\end{eqnarray}
for $U$-spin ladder operators,
\begin{eqnarray}
V_+ |\Xi^0\ra &=& |\Sigma^+\ra,\qquad\qquad\qquad\qquad\quad~ V_+ |\Sigma^-\ra=-|n\ra,  \nonumber\\
V_+|\Lambda\ra &=&-\frac{\sqrt{6}}{2} |p\ra,\qquad \qquad \qquad\qquad V_+ |\Sigma^0\ra=-\frac{\sqrt{2}}{2} |p\ra, \nonumber \\
 V_+|\Xi^-\ra &=& -\frac{\sqrt{2}}{2} |\Sigma^0\ra -\frac{\sqrt{6}}{2} |\Lambda\ra, \qquad\quad~
 V_+|\Xi_c^0\ra=|\Lambda_c^+\ra,  \\
V_-|p\ra &=& -\frac{\sqrt{2}}{2} |\Sigma^0\ra -\frac{\sqrt{6}}{2} |\Lambda\ra, \qquad\quad~
 V_-|\Sigma^+\ra=|\Xi^0\ra, \nonumber
\end{eqnarray}
for $V$-spin ladder operators, and
\begin{eqnarray}
I_+ |n\ra &=& |p\ra ,\qquad\qquad\qquad\qquad~~~~ I_+|\Xi^-\ra=|\Xi^0\ra,   \nonumber\\
I_+ |\Sigma^-\ra &=& \sqrt{2}|\Sigma^0\ra ,\qquad\qquad\qquad~~~~ I_+|\Sigma^0\ra=-\sqrt{2}|\Sigma^+\ra,   \\
I_+ |\Lambda\ra &=& 0,\qquad\qquad\qquad\qquad\qquad I_+|\Xi_c^0\ra=|\Xi_c^+\ra,   \nonumber\\
{{ I_- |\Xi_c^+\ra}} &=&{ |\Xi_c^0\ra,\qquad\qquad\qquad\qquad~~
{I_-|\Sigma^+ \ra= -\sqrt{2} |\Sigma^0 \ra}},  \nonumber
\end{eqnarray}
for isospin ladder operators.
Note some of the relations  may have signs different  from the textbook due to our wave function convention.
The ladder operators satisfy the commutator relations
\be
[U_+, U_-]=2U_3, \qquad [V_+, V_-]=2V_3, \qquad [I_+, I_-]=2I_3.
\en



\section{Form factors for $\Lambda_c^+$ decays}
\label{app:FF}
Form factors for $\Lambda_{c}^+\to \mathcal{B}$ transitions evaluated in the MIT bag model are shown in Table \ref{tab:FF1}. For $\Lambda_c^+\to p\eta_8$, we have assumed that form factors are dominated by the $(c\bar d)$ quark content.

\begin{table}[h]
\linespread{1.5}
\footnotesize{
 \caption{Same as Table \ref{tab:FF} except for $\Lambda_c^+$ decays.
 } \label{tab:FF1}
 \vspace{0.3cm}
\begin{center}
\begin{tabular}
{ lcccr|ccr}
 \hline\hline
 Modes & ($c\bar{q}$)&  $f_1(q_{\rm{max}}^2)$ &$ f_1(m_P^2)/f_1(q_{\rm{max}}^2)$ &$ f_1(m_P^2)$&  $g_1(q_{\rm{max}}^2)$ &
 $g_1(m_P^2)/g_1(q_{\rm{max}}^2)$ & $g_1(m_P^2)$  \\
 \hline
$\Lambda_{c}^{+}\to p \overline{K}^{0}$ & $(c\bar{d})$  & $-\frac{\sqrt{6}}{2}Y_1$  &$0.343423$ & $-0.371$  & $-\frac{\sqrt{6}}{2}Y_2$  &   $0.518518$ & $-0.488$  \\
$\Lambda_{c}^{+}\to \Lambda \pi^{+}$  & $(c\bar{s})$  & $Y_1^s$  &$0.440793$ & $0.419$  & $Y_2^s$  &$0.590594$  &$0.507$ \\
$\Lambda_{c}^{+}\to p \pi^{0}$  & $(c\bar{d})$  & $-\frac{\sqrt{6}}{2}Y_1$  &$0.305365$  &$-0.330$  & $-\frac{\sqrt{6}}{2}Y_2$  &$0.478571$  &$-0.450$ \\
$\Lambda_{c}^{+}\to p \eta $  & $(c\bar{d})$, $(c\bar{s})$  & $-\frac{\sqrt{6}}{2}Y_1$  &$0.353139$ & $-0.382$  & $-\frac{\sqrt{6}}{2}Y_2$  &$0.52837$  &$-0.497$ \\
$\Lambda_{c}^{+}\to n \pi^{+}$  & $(c\bar{d})$& $-\frac{\sqrt{6}}{2}Y_1$  &$0.306517$  &$-0.331$  & $-\frac{\sqrt{6}}{2}Y_2$  &$0.479606$  & $-0.451$\\
$\Lambda_{c}^{+}\to \Lambda K^{+}$  & $(c\bar{s})$& $Y_1^s$  &$0.494403$  &$0.470$  & $Y_2^s$  &$0.638728$ & $0.549$ \\
\hline\hline
\end{tabular}
\end{center}
 }
\end{table}

\section{Hadronic matrix elements and axial-vector form factors}
\label{app:me}

We use the MIT bag model to evaluate the baryon matrix elements and the axial-vector form factors (see e.g. \cite{Cheng:1991sn} for details).

\subsection{Baryon matrix elements}
\label{HME}

The  hadronic matrix elements $a_{\mathcal{B'\!B}}$ play an essential role both in $S$-wave
and $P$-wave amplitudes. The general expressions are given by
\begin{equation}
a_{\mathcal{B}'\!\mathcal{ B}} \equiv  \la \mathcal{B}'|\mathcal{H}_{\rm{eff}}^{\rm{PC}}|\mathcal{B}\ra
=
\left\{\begin{array}{ll}
\frac{G_F}{2\sqrt{2}} V_{cs} V^*_{ud} c_-\la \mathcal{B}' |O_- |\mathcal{B}\ra, & \textrm{CF}
\\ \\
\frac{G_F}{2\sqrt{2}} V_{cq} V^*_{uq} c_-\la \mathcal{B}' |O_-^q |\mathcal{B}\ra,
& \textrm{SCS}
\end{array}
\right.
\end{equation}
for CF and SCS processes, respectively, where $q=d,s$. Note in SCS process there are in general two operators.
For the definition of operators and Wilson coefficients, taking
CF process as an example, we have
 $O_-=(\bar{s}c)(\bar{u}d)-(\bar{s}d)(\bar{u}c)$, $c_-=c_1-c_2$ and then
we have the relation $c_+O_++c_-O_-=2(c_1O_1+c_2O_2)$.
The matrix element of $O_+$ vanishes since this operator is symmetric in color indices.
Below, we show the results of
 $\la \mathcal{B}' |O^{(q)}_- |\mathcal{B}\ra$ in the MIT bag model.

The relevant matrix elements for Cabibbo-favored processes are
\begin{align}
\label{eq:aCF}
&\la \Sigma^+|O_-| \Lambda_c^+ \ra= -\frac{2\sqrt{6}}{3}(X_1+3X_2) (4\pi), \quad
\la \Xi^0|O_-| \Xi_c^0\ra=\frac{2\sqrt{6}}{3}(X_1-3X_2)  (4\pi), \nonumber\\
&\la\Xi^0 |O_-|\Xi_c^{'0}\ra=-\frac{2\sqrt{2}}{3}(X_1+9X_2)  (4\pi), \quad~
\la\Sigma^0 |O_-|\Sigma_c^0\ra=-\frac{2\sqrt{2}}{3}(X_1-9X_2)   (4\pi),  \\
&{{\la\Sigma^+ |O_-|\Sigma_c^+\ra=\frac{2\sqrt{2}}{3} (X_1-9X_2) }}(4\pi),\quad\quad
\la\Lambda |O_-|\Sigma_c^0\ra=  -\frac{2\sqrt{6}}{3}(X_1+3X_2) (4\pi),  \nonumber
\end{align}
%
while the non-vanishing matrix elements for SCS decays are given by
\begin{align} \label{eq:aSCS}
&\la\Sigma^0|O_-^d|\Xi_c^{0}\ra= -\frac{4\sqrt{3}}{3}X_1^d (4\pi), \qquad\qquad
\la\Sigma^0|O_-^d|\Xi_c^{'0}\ra= \frac43X_1^d  (4\pi),\nonumber\\
&\la\Lambda|O_-^d|\Xi_c^{0}\ra= -4X_2^d (4\pi), \qquad\qquad\qquad
\la\Lambda|O_-^d|\Xi_c^{'0}\ra= -4\sqrt{3}X_2^d (4\pi), \nonumber\\
&\la p|O_-^d|\Sigma_c^{+}\ra= -\frac{2\sqrt{2}}{3}(X_1^d-9X_2^d) (4\pi), ~~
{{\la p |O_-^d|\Lambda_c^{+}\ra= \frac{2\sqrt{6}}{3}(X_1^d+3X_2^d)}}(4\pi),\nonumber\\
&\la n|O_-^d|\Sigma_c^{0}\ra= \frac43(X_1^d+9X_2^d) (4\pi), \qquad\quad
 \la\Sigma^0|O_-^s|\Xi_c^{'0}\ra= -\frac23(X_1^s-9X_2^s) (4\pi),  \nonumber\\
&\la\Lambda|O_-^s|\Xi_c^{0}\ra= -2(X_1^s-X_2^s) (4\pi),  \qquad\quad
\la\Lambda|O_-^s|\Xi_c^{'0}\ra= -\frac{2\sqrt{3}}{3} (X_1^s+ 3 X_2^s)(4\pi), \\
&\la \Xi^0 |O_-^s|\Omega_c^{0}\ra= -\frac43(X_1^s+9X_2^s) (4\pi),   \qquad
\la\Sigma^+|O_-^s|\Xi_c^+\ra= \frac{2\sqrt{6}}{3}(X^s_1+3X^s_2) (4\pi),  \nonumber\\
&\la\Sigma^+|O_-^s|\Xi_c^{'+}\ra=\frac{2\sqrt{2}}{3}(X_1^s-9X_2^s)  (4\pi), \quad
\la\Sigma^0|O_-^s|\Xi_c^{0}\ra= -\frac{2\sqrt{3}}{3}  (X_1^s + 3X_2^s) (4\pi), \nonumber
\end{align}
where we have introduced the bag integrals
\begin{align}
& X_1= \int^R_0 r^2 dr (u_s v_u -v_s u_u)(u_c v_d -v_c u_d), \quad
X_2=\int^R_0 r^2 dr(u_s u_u+ v_s v_u) (u_c u_d + v_c v_d),\nonumber\\
&X_1^q=\int^R_0 r^2 dr (u_q v_u-v_q u_u)(u_q v_c -v_q u_c), \quad
X_2^q=\int^R_0 r^2 dr (u_q v_u+v_q u_u)(u_q v_c +v_q u_c),  \nonumber\\
\end{align}
with $q=d, s$. Numerically, we obtain 
\begin{align} \label{eq:Xi}
& X_1^d=0, \qquad X_2^d=1.60\times 10^{-4}, \quad X_1^s=2.60\times 10^{-6},\nonumber \\
& X_2^s=1.96\times 10^{-4}, \quad
X_1=3.56\times 10^{-6}, \quad X_2=1.74\times 10^{-4}. 
\end{align}

\subsection{Axial-vector  form factors}
\label{sec:axial-vectorFF}
In the MIT bag model the axial form factor in the static limit can be expressed as
\begin{equation}
g^{A(P)}_{\mathcal{B}'\mathcal{B}}=\la \mathcal{B}'\uparrow | b_{q_1}^\dagger b_{q_2}\sigma_z |
\mathcal{B}\uparrow\ra \int d^3\bm{r}\left(u_{q_1}u_{q_2}-\frac13 v_{q_1}v_{q_2}\right).
\label{eq:gA}
\end{equation}
Based on Eq. (\ref{eq:gA}),
the axial-vector form factors related to CF processes are
\footnote{Recall that the axial-vector current  is $(\bar u\gamma_\mu\gamma_5 u-\bar d\gamma_\mu\gamma_5 d)/2$ for $\pi^0$ and $(\bar u\gamma_\mu\gamma_5 u+\bar d\gamma_\mu\gamma_5 d-2\bar s\gamma_\mu\gamma_5 s)/(2\sqrt{3})$ for $\eta_8$ in our convention.
}
\begin{align}
 g^{A(\overline{K}^0)}_{\Lambda_c^+ \Xi_c^+}
& =g^{A(\pi^+)}_{\Xi_c^0 \Xi_c^+}= g^{A(\pi^0)}_{\Xi_c^0 \Xi_c^0} = g^{A(\eta_8)}_{\Xi_c^0 \Xi_c^0}=0, \\
 g^{A(\pi^+)}_{\Xi_c^{'0}\Xi_c^+} &=\sqrt{3} g^{A(\pi^+)}_{\Xi^- \Xi^0}=-2 g^{A(\pi^0)}_{\Xi_c^{'0} \Xi_c^0}
=2\sqrt{3} g^{A(\pi^0)}_{\Xi^0 \Xi^0}
=\frac{2}{\sqrt{3}}g^{A(\eta_8)}_{\Xi_c^{'0} \Xi_c^0}  \nonumber \\
& =\frac{2}{3} g^{A(\eta_8)}_{\Xi^0 \Xi^0}
=g^{A(\eta_8)}_{\Lambda\Lambda}=
-\frac{\sqrt{3}}{3}(4\pi Z_1),   \\
 g^{A(\overline{K}^0)}_{\Sigma_c^+ \Xi_c^+} &=\frac{\sqrt{2}}{2} g^{A(\overline{K}^0)}_{\Sigma_c^0 \Xi_c^0}
=\sqrt{2} g^{A(\overline{K}^0)}_{\Lambda \Xi^0}
=-\frac{\sqrt{6}}{5} g^{A(\overline{K}^0)}_{\Sigma^0 \Xi^0}
=\frac{\sqrt{3}}{5} g^{A(K^-)}_{\Sigma^+ \Xi^0}
=\frac{\sqrt{3}}{3}(4\pi Z_2),
\end{align}
and
\begin{align}
g^{A(\eta_8)}_{\Xi_c^+ \Xi_c^+} & =  g^{A(\eta_8)}_{\Sigma^0 \Lambda}
=g^{A(\eta_8)}_{ \Lambda \Sigma^0}
=g^{A(\pi^0)}_{\Lambda\Lambda}
=g^{A(\pi^0)}_{\Sigma^0\Sigma^0}=0,  \\
g^{A(\pi^0)}_{\Lambda\Sigma^0} & = g^{A(\pi^0)}_{\Sigma^0\Lambda}=g^{A(\eta_8)}_{\Sigma^0\Sigma^0}
=g^{A(\eta_8)}_{\Sigma^+\Sigma^+}
={{-}} g^{A(\eta_8)}_{\Lambda\Lambda}=-\frac{2\sqrt{3}}{3}g^{A(\eta_8)}_{\Xi_c^{'+}\Xi_c^+} \nonumber \\
& =\frac{\sqrt{3}}{2}g^{A(\pi^0)}_{\Sigma^+\Sigma^+}=\frac{\sqrt{2}}{2} g^{A(\pi^+)}_{\Lambda\Sigma^+},
 = -\frac{\sqrt{6}}{4} g^{A(\pi^+)}_{\Sigma^0 \Sigma^+} =\frac{\sqrt{6}}{4} g^{A(\pi^+)}_{\Sigma^- \Sigma^0}   \\
& =\frac{\sqrt{2}}{2} g^{A(\pi^+)}_{\Sigma^- \Lambda} =-\frac{\sqrt{6}}{4} g^{A(\pi^-)}_{\Sigma^+ \Sigma^0}
=\frac{\sqrt{2}}{2} g^{A(\pi^-)}_{\Sigma^+ \Lambda}=\frac{\sqrt{3}}{3}(4\pi Z_1),
\nonumber\\
g^{A(\overline{K}^0)}_{p\Sigma^+}&= -\frac{\sqrt{6}}{6}g^{A(K^+)}_{\Omega_c^0 \Xi_c^+}
= -\frac{\sqrt{6}}{6}g^{A(K^0)}_{\Omega_c^0 \Xi_c^0}=-\frac{\sqrt{2}}{5} g^{A(K^0)}_{\Xi^0 \Sigma^0}
=\frac{\sqrt{6}}{3}g^{A(K^0)}_{\Xi^0 \Lambda} \nonumber \\
& =-\frac{\sqrt{2}}{5} g^{A(K^+)}_{\Xi^- \Sigma^0}
=-\frac{\sqrt{6}}{3}g^{A(K^+)}_{\Xi^- \Lambda}
= \sqrt{2} g^{A(K^-)}_{p\Sigma^0} = -\frac{\sqrt{6}}{9} g^{A(K^-)}_{p\Lambda} \\
& = -\frac{\sqrt{6}}{9} g^{A(\overline{K}^0)}_{n\Lambda}
=-\sqrt{2} g^{A(\overline{K}^0)}_{n\Sigma^0}
{{=\frac15 g^{A(K^+)}_{\Xi^0 \Sigma^+}}}
=\frac13 (4\pi Z_2), \nonumber
\end{align}
for SCS processes,
where the auxiliary parameters are introduced
\begin{equation}
Z_1=\int r^2 dr\left(u_u^2 -\frac13 v_u^2\right),\quad
Z_2=\int r^2 dr \left(u_u u_s -\frac13 v_u v_s\right)
\end{equation}
in the bag model. The numerical results are
\begin{equation} \label{eq:Zi}
(4\pi) Z_1= 0.65\,, \qquad (4\pi)Z_2=0.71\,.
\end{equation}



\end{document}